# The First Instrumentally Documented Fall of an Iron Meteorite: atmospheric trajectory and ground impact

Jarmo Moilanen[1,2], Maria Gritsevich[2,3] , Jaakko Visuri[2]

Abstract

Iron meteorite falls are rare compared to stony meteorites, and until recently no iron meteorite has had a reliably determined pre-atmospheric orbit. This changed on 2020 November 7, when a bright fireball was observed across Sweden and neighboring regions, with optical, acoustic, and seismic detections extending up to 665 km from the trajectory. After a month-long recovery effort, a 13.8 kg iron meteorite was discovered near Ådalen, representing the first instrumentally recorded and recovered fall of its type and the first iron meteorite with a derivable heliocentric orbit; the event also exhibited the lowest terminal height measured for a well-documented fireball. We combine optical, infrasound, and seismic data to reconstruct the luminous trajectory and employ a Monte Carlo model to simulate the dark flight phase and predicted strewn field, while also investigating the plausibility of a ricochet prior to final deposition. Our analysis identifies distinct aerodynamic properties of iron meteoroids compared to stony bodies, including the influence of streamlined shapes and deep regmaglypts on drag and flight stability, underscoring the need to incorporate iron-specific parameters into entry models to constrain atmospheric dynamics and improve recovery predictions for future events.

[1] Corresponding author jarmo.moilanen@somerikko.net
[2] Finnish Fireball Network, Ursa Astronomical Association, Helsinki, Finland
[3] Instituto de Astrofísica de Andalucía (IAA-CSIC), Glorieta de la Astronomía s/n, E-18008, Granada, Spain

# 1 Introduction

Meteorites play an essential role in unravelling the conditions of planetary systems and celestial bodies formation, including that of Earth. They provide access to elemental and isotopic compositions that are otherwise challenging, if not impossible, to obtain on our planet. Serving as a cornerstone of cosmochemistry, meteorites offer insights into the intricate processes that shaped our solar system.

Among celestial remnants, iron meteorites stand out for their exceptional scientific significance, aesthetic appeal, and collectible nature. Studies of iron meteorites aid in comprehending the complex formation processes of their parent bodies, with a particular focus on deciphering their thermal history, crystallization timelines, cosmic-ray exposure ages, and, more recently, identifying potential iron-rich reservoirs within the Solar System (J. Goldstein et al. 2009; E. Kaminski et al. 2020; E. R. D. Scott 2020; I. Kyrylenko et al. 2023).

Despite their relative rarity, constituting a mere 2% of all officially classified meteorites by numbers (Table 1), iron meteorites are distinguished by their recognizable appearance and durability, often surviving the descent through Earth's atmosphere nearly intact. With their high metal content, iron meteorites tend to be more resilient than stony ones (M. Gritsevich et al. 2024), representing most of the meteoritic mass identified on Earth's surface, as well as on Mars. Composed primarily of iron (Fe) and nickel (Ni), with traces of elements like cobalt and other trace elements, these objects are believed to be remnants of once-molten metallic cores or magma chambers within planetesimals that underwent a geological process known as differentiation driven by heat and gravity. This process involved the separation of metal from silicate materials, giving rise to the formation of metallic cores within their parent bodies.

Meteorites with known orbits are particularly invaluable as they establish a direct link between the meteorite and its source region in space, often an asteroid. Orbit calculations for meteorites provide knowledge about their parent bodies, and potential associations with other meteorites or asteroid families (V. Dmitriev et al. 2015; J. M. Trigo-Rodríguez et al. 2015; M. Granvik & P. Brown 2018; T. Jansen-Sturgeon et al. 2019; E. Peña-Asensio et al. 2021). This is significant to our understanding of the dynamic processes within the solar system and the origins of meteorites.

As of July 21, 2026, iron meteorite falls (i.e., those observed during their descent through the atmosphere) remain exceedingly rare, comprising just 3.98% of the 1280 known witnessed meteorite falls with approved names in MBD (2026). None of the 49 previously confirmed and documented iron meteorite falls were sufficiently recorded with instrumentation to allow for precise orbit calculations (Table 1). The broader population of iron meteoroids in near-Earth space is also difficult to characterize due to the challenges of constraining composition from remote sensing alone and the paucity of well-documented events. Some studies infer their presence from fireball and meteor spectra, ablation and luminous efficiency models, and deceleration profiles, which in certain cases are interpreted as signatures of high-density, iron objects (E. Peña-Asensio & M. Gritsevich 2025). However, conclusions drawn from different models diverge significantly, reflecting both methodological uncertainties and the limited observational dataset. Consequently, prior to the Swedish meteorite fall on November 7, 2020, which is the focus of this study, determining the Solar System orbit for a confirmed iron meteoroid was not possible.

Table 1. Statistics of iron meteorites and iron-rich asteroids. Only asteroids classified as M-type in the Tholen taxonomic system are included as iron asteroids. A) Upper limit estimate. B) Only meteorites with approved names in the Meteoritical Bulletin Database (MBD, 2026) are included. C) Includes both the Swedish (2020) and the Polish (2026) instrumentally documented iron meteorite falls.

| Objects | All | Irons | Irons % | Source |
|---|---|---|---|---|
| Known asteroids | 1 553 113 | 38 – 95 [A] | 0.003 – 0.006 [A] | 1, 2 |
| With known spectrum | 623 827 | 38 | 0.006 | 2 |
| Registered meteorites [B] | 80 074 | 1440 | 1.80 | 3 |
| Meteorite falls [C] | 1280 | 51 | 3.98 | 3, 4, 5, 6, 7 |
| Falls with orbit [C] | 90 | 2 | 2.22 | 7 |

References: 1: MPC 2026, 2: SSD 2023, 3: MBD 2026, 4: I. Kyrylenko et al. 2023, 5: P. Jenniskens & H. A. R. Devillepoix (2025), 6: J. Borovička et al. 2026, 7: P. Jenniskens (2026).

In the early 20th century, efforts to determine the trajectories and orbits of iron meteorites faced considerable challenges due to limited data availability (Pickering, 1910; I. S. Astapovich, 1939). However, on February 12, 1947, a remarkable daytime bolide brighter than the sun streaked across the sky over the Sikhote-Alin Mountains in Primorye, Soviet Union. Eyewitnesses observed its descent at an angle of about 41 degrees, visible over a vast area extending 300 kilometers from the impact point near Luchegorsk. Upon entering the atmosphere, the iron asteroid heavily fragmented, with some fragments burying themselves up to 6 meters deep upon impact (E. L. Krinov 1966; V. F. Buchwald 1975, pages 1123-30).

This widely witnessed event allowed for extensive analysis of reported data. V. G. Fesenkov, chairman of the meteorite committee of the USSR Academy of Science at the time, utilized these observations to reconstruct the atmospheric trajectory and estimate the pre-impact orbit before encountering Earth. These findings were used to determine key parameters such as the ballistic coefficient and mass loss parameter, essential for putting the event in larger context (V. P. Stulov 2006; M. Gritsevich et al., 2009, 2012; L. I. Turchak & M. Gritsevich, 2014; T. Stevenson et al., 2026).

Recent decades have seen significant technological advancements, especially in the use of casual surveillance videos and photographic registrations, which have inadvertently also led to the global collection of data on occasionally registered meteorite falls. Despite the fast increase of recording devices including dedicated fireball camera systems (A. J. Castro-Tirado, 2023, P. Jenniskens & H. A. R. Devillepoix, 2025), cell phones and surveillance cameras, none of notable iron specimens fall recently, such as Sokoto (2008, Nigeria), Kavarpura (2006, India), Ban Rong Du (1993, Thailand), and Sterlitamak (1990, Russia), were instrumentally documented (I. Kyrylenko et al. 2023).

Only the most recent documented iron meteorite fall, Zadzim (prov.) in Poland, on 17 April 2026, has yielded a reconstructed orbit based on observations from the Skytinel network in Poland and the European Fireball Network in the Czech Republic (J. Borovička et al., 2026). Including the Zadzim fall in current meteorite statistics, approximately 3.9% of recovered iron meteorites now have known pre-impact orbits, compared with about 6.9% of recovered non-iron meteorites (Meteoritical Bulletin Database, 2026; Jenniskens, 2026).

The orbit and possible origin of the Swedish meteorite were recently elucidated in a companion study by I. Kyrylenko et al. (2023). The Norwegian meteor network has also published preliminary orbital parameters on a page related to the fireball[4]. In this paper, we present, interpret, and make available all observations of the iron meteorite fall in Ådalen, Sweden, on November 7, 2020. This includes sound data analysis, the reconstructed trajectory, and detailed DFMC simulations of the strewn field, along with a morphological description of the recovered meteorite. We also provide a comprehensive scenario for its ground impact, as exceptionally it is available with a high level of detail for the Swedish iron meteorite case. Additional observational material, representative camera images, measurement uncertainties, and the partial three-dimensional shape model of the recovered meteorite are provided in the Supplementary Material.

## 2 The Fireball and the Meteorite Recovery

On November 7, 2020, at 21:27:04 UTC (22:27:04 local time), an exceptionally bright fireball illuminated the skies over Sweden for approximately 4.2 seconds. The lowest point where the fireball was instrumentally documented was 11.28 km, marking the event as the deepest atmospheric penetration ever recorded for a fireball (J. Moilanen & M. Gritsevich 2021; 2022; I. Kyrylenko et al. 2023). Detailed instrumental records were acquired by the Finnish Fireball Network (Fig. 1) and the Norwegian meteor camera network, as well as several cameras in Denmark. Unfortunately, most fireball cameras in Sweden were obscured by clouds, capturing only the illumination scattered by the clouds. A surveillance camera recording from a private house located directly beneath the fireball trajectory was similarly affected. The locations of video observations are shown in Figure 2. See Figure 1 in I. Kyrylenko et al. (2023) for additional details, including the orbital elements derived from the fireball.

In Finland, 28 observations, including seven photographic or video records (Table 2), were submitted via the Ursa Astronomical Association's Taivaanvahti ("Skywarden" in English) observation service (Ursa 2022)[5]. Additionally, 22 eyewitness reports were gathered from Denmark, Norway, and Sweden by the International Meteor Organization[6]. An interesting example is a report by "Max B", a flight captain on Frankfurt to Stockholm flight LH808. He reported witnessing the fireball while descending through an altitude of 23,470 feet (7,153 m), stating: “I definitely saw it below the aircraft when it disappeared. The sky was as bright as during daylight, even the sky was blue for a few seconds”.

[4] https://norskmeteornettverk.no/meteor/20201107/212700/
[5] https://www.taivaanvahti.fi/
[6] https://fireball.imo.net/members/imo_view/event/2020/6403

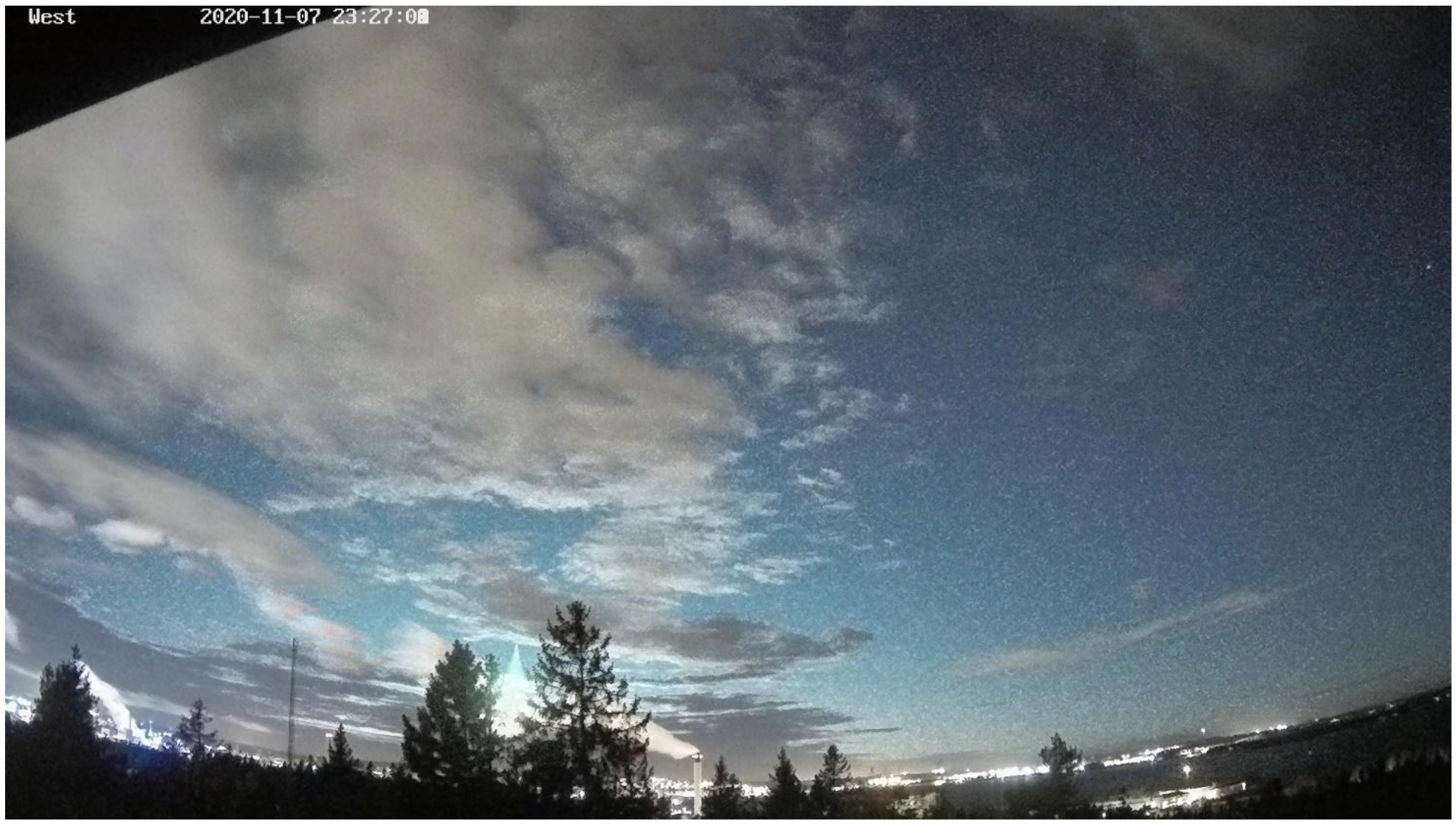

Figure 1. Composite image constructed from all video frames recorded the Finnish Fireball Network at Tampere, Finland, showing the fireball illuminating the night sky. The fireball is visible between the two tallest trees. Image © Markku Lintinen, Tampereen Ursa ry.

NASA also registered the fireball, including it in the CNEOS fireball and bolide database[7] (E. Peña-Asensio et al. 2025). According to the CNEOS database, the peak brightness occurred at 21:27:04 UT, with coordinates indicating a latitude of 59.8° N and longitude of 16.8° E. The peak brightness altitude was 22.3 km, with a projectile velocity of 16.7 km/s. The calculated total impact energy of the event was 0.33 kilotons of TNT equivalent (equivalent to 1.38 terajoules). CNEOS data also provides orbital elements for this fireball.

[7] https://cneos.jpl.nasa.gov/fireballs/

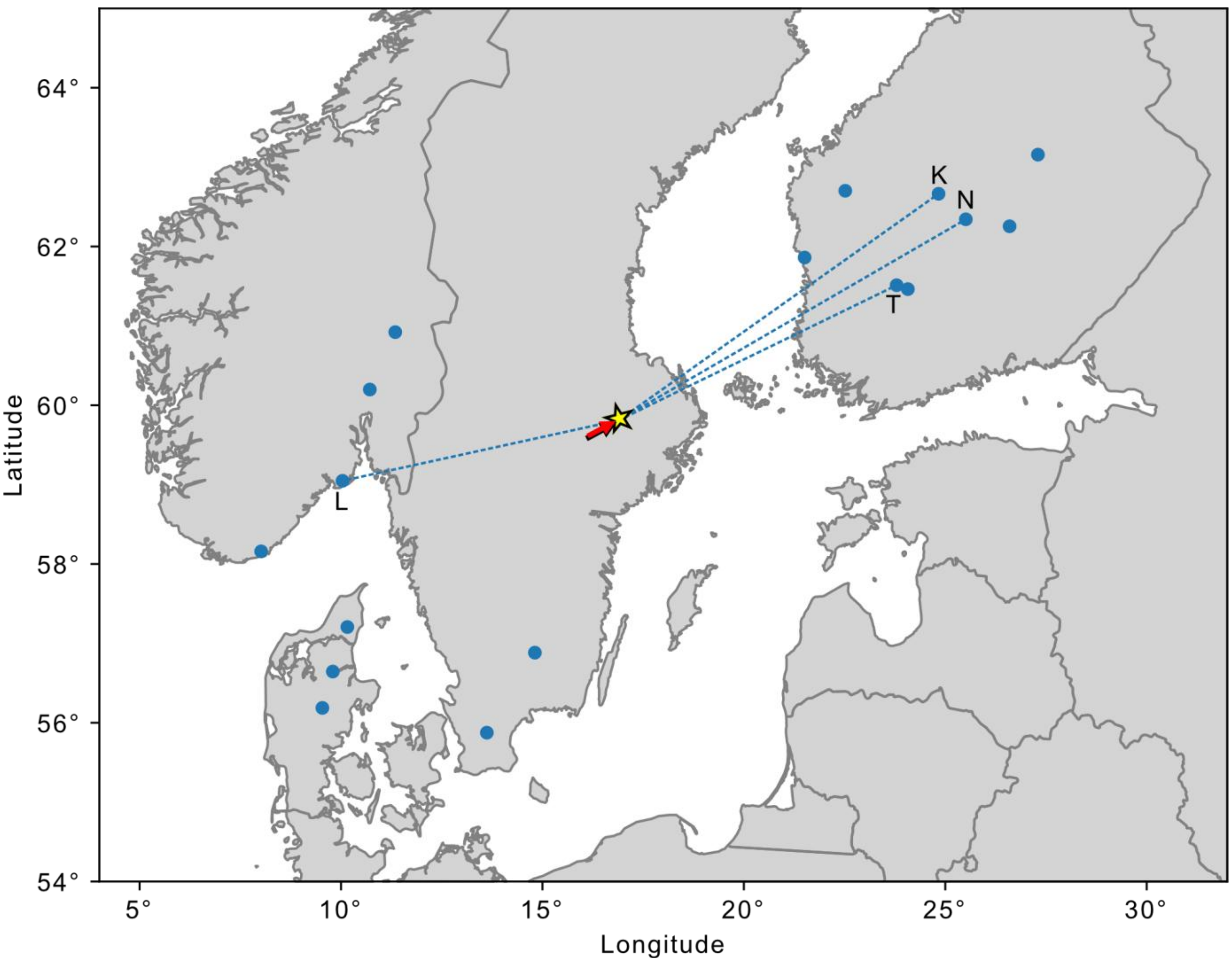

Figure 2. Solid circles show video observations of the 7 November 2020 fireball across Nordic countries. The star marks the meteorite fall site near Enköping, Sweden. The arrow shows the projected ground track of the fireball trajectory. Observations used for triangulation are labeled as T = Tampere, N = Nyrölä observatory, and K = Kyyjärvi (all in Finland), and L = Larvik (Norway).

Table 2. The 28 Finnish observations of the Swedish fireball reported in Taivaanvahti.fi observation service. Individual observations can be accessed by appending the corresponding observation ID to the URL. For example, observation ID 94754 is available at: https://taivaanvahti.fi/observations/show/94754.

| ID | Observation | Location | Observer | Time [+2h UT] | Distance [km] |
|---|---|---|---|---|---|
| 94754 | surv. video | Merikarvia | Helinkö H. | 23:26 | 335 |
| 94749 | sky cam video | Tampere obs., Tampereen Ursa | Kuure K. A. & Lintinen M. | 23:27 | 420 |
| 94839 | surv. video | Kangasala | Pyykkö J. | 23:27 | 435 |
| 94929 | dashcam video | Ilmajoki | Lahola V. | 23:22 | 440 |
| 94766 | all-sky photo | Nyrölä obs. Jyväskylän Sirius | Kiiskinen H. | 23:27 | 540 |
| 94767 | all-sky photo | Hankasalmi obs. Jyväskylän Sirius | Oksanen A. | 23:26 | 590 |
| 94802 | *surv. video | Maaninka | anon. | 23:27 | 665 |
| 94759 | visual | Laitila | Koski P. | 23:27 | 285 |
| 94739 | visual | Kemiönsaari | Odé J. | 23:19 | 310 |
| 94752 | visual | Turku | Nieminen O. | 23:27 | 310 |
| 94800 | visual | Eura | anon. | 23:26 | 320 |
| 94765 | visual | Pori | Pohjola T. | 23:28 | 325 |
| 94760 | visual | Merikarvia | Rosendahl M. | 23:30 | 330 |
| 94776 | visual | Merikarvia | Seranto P. | 23:25 | 335 |
| 94741 | visual | Karijoki | Malmelöv R. | 23:27 | 375 |
| 94924 | visual | Sastamala | Hesso A. & Seppä V. | 23:26 | 375 |
| 94756 | visual | Hämeenkyrö | anon. | 23:45 | 395 |
| 94794 | visual | Kauhajoki | Huhtanen J. | 23:30 | 405 |
| 94827 | visual | Körsnäs | Höstman P. | 23:30 | 410 |
| 94778 | visual | Ylöjärvi | anon. | 23:40 | 410 |
| 94780 | visual | Ikaalinen | Rikala H. | 23:30 | 420 |
| 94740 | visual | Maalahti | Olmari J. | 23:28 | 420 |
| 94792 | visual | Ylöjärvi | anon. | 23:00 | 430 |
| 94786 | visual | Ilmajoki | Niemelä J. | 23:28 | 445 |
| 94742 | visual | Ruovesi | anon. | 23:10 | 455 |
| 94785 | visual | Vöyri | Länne S. | 23:30 | 480 |
| 94809 | visual | Muurame | anon. | 23:00 | 540 |
| 94738 | visual | Vihanti | Saarivainio H. | 23:26 | 665 |

Notes: Timing of visual observation is often inaccurate. Distances to a point in the fall area are given in ±5 km accuracy because coordinates for some observation sites are not accurate and distance to the observed part of the fireball differs. The ID column lists the corresponding observation number in the Taivaanvahti.fi database. *) This video is not available through the Taivaanvahti.fi observation service.

The observed low terminal altitude of the fireball made it exceptional among instrumentally documented meteor events (M. Moreno-Ibáñez et al. 2015; E. K. Sansom et al. 2019). This deep luminous entry alone hinted at the possibility of an iron meteorite fall. Furthermore, the hypothesis of an iron meteorite origin was bolstered by the fact that a surveillance video showed that sonic booms were heard just 29 seconds after the fireball's termination. This timing aligns with the 11.28 km terminal altitude (I. Kyrylenko et al. 2023) and a steep trajectory angle of 72°.

Such deep penetration is only possible for a relatively heavy meteorite, implying a considerable size (possibly hundreds of kilos), a low aerodynamic coefficient, or a dense meteoritic material. The upper known bound for meteorite material corresponds to iron meteorites with bulk densities ranging between 7 to 8 g/cm$^3$ (Ostrowski & Bryson 2019). Consequently, following the strewn field determination model (DFMC) described by J. Moilanen et al. (2021), we decided to run simulations using an iron meteorite density of 7.4 ± 0.5 g/cm$^3$.

Our hypothesis gained further support when, in January 2021, it was publicly disclosed that a few tiny fragments of possible iron meteorite fusion crust have been found on November 20, 2020 (M. Testorf 2021a). These fragments were discovered near a large boulder exhibiting an impact mark, possibly left by a sizable meteorite fragment (Fig. 8). On February 23, 2021, the discovery of a 13.8 kg iron meteorite was announced (M. Testorf 2021b). This meteorite was found atop a bedrock outcrop (Fig. 10) in early December, approximately 75 meters from the boulder bearing the impact mark.

X-ray fluorescence (XRF) composition analysis of small surface fragments and photographs of the actual meteorite (Fig. 7a), as published in M. Testorf (2021a, b), left no room for doubt—it was an iron meteorite. Unfortunately, as of the Meteoritical Bulletin database (MBD 2026), this meteorite has not yet received official classification. The main mass is housed in the Natural History Museum of Sweden in Stockholm, though legal issues surrounding its ownership have hindered the study and classification process (Radio Sweden 2022). Presently, the meteorite still lacks an official name; however, it has been referred to as Ådalen after the nearest village (I. Kyrylenko et al. 2023; P. Jenniskens & H. A. R. Devillepoix (2025); J. Visuri et al. 2026). On 19 August 2025 the Supreme Court of Sweden issued a final ruling on ownership, determining that the meteorite belongs to its finders.[8] It is hoped that this decision will facilitate the completion of the official classification process for this remarkable meteorite specimen.

## 3 Observations

The Ådalen fireball was exceptionally bright and widely observed across Finland, Norway, and Denmark, with reports and recordings from numerous locations, even those far from the peak brightness area. However, Sweden was heavily cloud-covered at the time, limiting observations from that region.

---

[8] https://www.domstol.se/hogsta-domstolen/nyheter/2025/08/meteoriten-tillhor-upphittarna/

A Finnish internet observation service, Taivaanvahti, managed by the Ursa Astronomical Association, received 28 reports of the fireball (Table 2). Observations spanned a wide area in southern Finland, extending as far as Vihanti and Maaninka, approximately 665 km from the fall site. Under clear skies and without horizon obstructions, the fireball could have been visible across an area of 1.3 million square kilometers (1/400 of Earth's surface). In J. Moilanen & M. Gritsevich (2021), observations from Kempele and Kuopio (~720 km from the fall site) were initially noted, but further analysis suggested that these likely corresponded to a separate fireball shortly after the Swedish event.

At least seven cameras in Finland captured the Ådalen fireball: four from the Finnish Fireball Network (M. Gritsevich et al., 2014b), two from private security systems, and one from a dashboard camera. However, Finnish cameras did not record the fireball's endpoint due to its exceptionally low terminal altitude and horizon obstructions (I. Kyrylenko et al., 2023).

### 3.1 Trajectory Triangulation

Despite many recordings being obstructed by trees or background objects due to the fireball's low elevation, four key observations — three from the Finnish Fireball Network and one from the Norwegian Meteor Network — were particularly useful for triangulating its trajectory. These observations, recorded from Kyyjärvi, Nyrölä, and Tampere (Finland) and Larvik (Norway) (Table 3, 4, and 5, see also Supplementary Material), were used to reconstruct the trajectory with FireOwl, the latest software developed for processing data acquired by the Finnish Fireball Network (J. Visuri & M. Gritsevich, 2021; I. Kyrylenko et al., 2023; M. Gritsevich et al. 2024; E. Peña-Asensio et al., 2024; J. Visuri et al. 2026). According to I. Kyrylenko et al. (2023), the beginning point of the luminous trajectory was at latitude 59.74° and longitude 16.51°, at a height of 81.53 km, with an initial velocity of 17.52 km $s^{-1}$. The terminal point was located at latitude 59.82° and longitude 16.84°, at a height of 11.28 km.

The Larvik camera in Norway captured high-quality video at 15 fps, recording the fireball's lowest observed altitude. However, despite being the closest camera to the entry location, it did not capture the terminal part of the fireball. After applying the δz correction (J. Visuri et al. 2026), the minimum observed altitude was refined to 11.42 km. The lowest point of the visible trajectory was determined from the Larvik station, which had a clear view of the horizon. None of the meteor cameras recorded the final endpoint of the luminous flight.

The Kyyjärvi camera in Finland was the outermost used for triangulation. Due to hardware and processing incompatibility, its x/y ratio is not 1:1 (see Table 5). As an all-sky camera, it did not capture the fireball at low elevations but provided high-quality azimuthal direction data.

The Nyrölä camera, also in Finland, was another all-sky camera that provided azimuthal direction data. Although the fireball was partially obscured by tree branches, it remained visible since it was not blocked by trunks. The fireball's trajectory was nearly perpendicular to the horizon, as its ground projection moved toward the camera.

The Tampere camera, observing from a similar direction as Nyrölä, was affected by poor weather conditions that prevented capturing the fireball at low elevations. The video frame rate (~3 fps) allowed for velocity and directional data collection only during the initial phase of the flight.

The limiting factor for the trajectory reconstruction was the observational geometry rather than detector sensitivity. At all stations used for triangulation, the fireball brightness exceeded the detector noise by a large margin. Consequently, the accuracy of the reconstructed trajectory was governed primarily by camera calibration, image resolution, and partial horizon obstruction.

In addition to the camera-calibration residuals listed in Table 4, the Supplementary Material provides the angular image scale for each observing station, the estimated angular measurement uncertainty, and the number of measured fireball positions used in the trajectory solution. These quantities characterize the angular precision of the measurements employed in the trajectory reconstruction. The combined slope-and-azimuth orientation of the reconstructed trajectory has an estimated uncertainty of 0.303°.

The trajectory calculations also incorporated the atmospheric refraction δz-correction technique (J. Visuri et al., 2026). Owing to the viewing geometry of the Ådalen fireball, the δz-correction has a negligible effect on the reconstructed trajectory and the derived heliocentric orbit. However, it significantly affects the measured velocities at lower altitudes, reducing the terminal velocity by approximately 560 m $s^{-1}$ (J. Visuri et al., 2026). Applying the δz-correction was therefore essential for accurately determining the meteoroid's deceleration, terminal mass, ground impact conditions, and the predicted strewn field, even though its influence on the reconstructed orbit is minimal.

The reconstructed atmospheric trajectory was independently evaluated against the recorded acoustic observations. Sound propagation was calculated using atmospheric temperature profiles representative of the event. The calculated sonic-boom arrival times are consistent with the observed acoustic signals within the expected uncertainties, providing an independent validation of the reconstructed trajectory.

Table 3. Location of camera stations used for trajectory calculations of the fireball.

| Station | Latitude [N°] | Longitude [E°] | Elevation [m] |
|---|---|---|---|
| Kyyjärvi, Finland | 63.076695 | 24.749890 | 244 |
| Nyrölä, Finland | 62.342593 | 25.509586 | 203 |
| Tampere, Finland | 61.511861 | 23.793079 | 171 |
| Larvik, Norway | 59.090907 | 10.098390 | 93 |

Table 4. Optical and projection parameters for the stations used in triangulation, derived following the camera calibration and trajectory-fitting method described in M. Gritsevich et al. (2024). The third column lists the root-mean-square values of the star-fitting residuals.

| Station | stars used | res. (RMS) [pix] | Azimuth [°] | Elevation [°] | Center X [pix] | Center Y [pix] | Tilt [°] |
|---|---|---|---|---|---|---|---|
| Tampere | 21 | 0.767 | 260.946 | 21.623 | 985.537 | 1990.582 | 0.6045 |
| Nyrölä | 25 | 0.370 | 104.826 | 85.464 | 506.163 | 299.986 | -84.3642 |
| Kyyjärvi | 16 | 0.714 | -278.436 | 89.548 | 361.323 | 266.996 | -83.4469 |
| Larvik | 16 | 0.625 | 239.35302 | 133.79481 | 1031.83182 | 692.85280 | -179.0776 |

Table 5. Projection variables for the camera stations used in the trajectory calculations of the fireball.

| Station | x/y ratio | f [pix] | t | k_5 | k_7 |
|---|---|---|---|---|---|
| Tampere | 1 | 1287.158 | 1.485 | -0.0428 | 0.0954 |
| Nyrölä | 1 | 247.37 | 1.0256 | -0.0013 | 0.0007 |
| Kyyjärvi | 0.9374 | 191.042 | 1.129 | 0.0002 | -0.0002 |
| Larvik | 1 | 1229.992 | 1.4073 | -0.0195 | 0.0377 |

As a result of the trajectory determination and following our standard criteria for selecting the DFMC starting point (see discussion in M. Gritsevich et al. 2024), the DFMC simulations (Section 4.1) were initiated from latitude 59.78° and longitude 16.69°, at a height of 43.89 km, with a velocity of 17.34 km $s^{-1}$. The adopted default uncertainties are ±0.3 km for position and height, and ±0.3 km $s^{-1}$ for velocity (J. Moilanen et al. 2021).

## 3.2 The Surveillance Video

An additional information source used in investigating the later stages of meteorite fall was a home security video recorded in Björkbacken (WGS84 coordinates: 59.7845° N, 16.8243° E), also known as Svinn, a nearby residential area. This location is situated just over six kilometers southwest of the fall site. The video captured flashes of light outside the house, followed by loud booms and rumbling sounds. Due to low-light conditions, the recording has a frame rate of 10 frames per second, introducing some timing uncertainties in the image frames. Additionally, since the camera was positioned inside the house, the audio quality was affected. A version of the video can be viewed on YouTube[9].

Since the fireball itself is not visible in this video, we used alternative methods to produce a light curve. The average color (RGB) value curve was determined by extracting pixel values from six locations outside the house. Also, one dark reference spot was chosen inside the house. In Figure 3, the upper curve represents the average brightness outside the house, while the lower curve depicts the dark reference. The time 0.0 corresponds to the peak brightness of the fireball. This moment is assumed to match the timing 2020-11-07 21:27:04 UT reported in NASA's CNEOS database, since the security camera footage did not contain time stamps. The light curve in Figure 3 covers 2.8-7.1 seconds of the recording, with timestamp 0.0 occurring 4.9 seconds after the start of the video. The last detectable trace of light from the fireball appears at 5.9 seconds from the beginning of the footage (corresponding to 1.0 seconds in Figure 3).

[9] https://www.youtube.com/watch?v=XwY1Wib5d5U

Both curves reflect the impact of the camera's auto-exposure function, which took more than nine seconds to return to levels prior to the fireball. The light curve also reveals a subtle 0.3-second pulse in brightness occurring at -1.8, -1.5, -1.2, and -0.9 seconds. While this may be an artifact, it could also indicate a 0.6-second periodicity in the rotation of the entering mass. We did not pursue further investigation of this hypothesis, but a dedicated light curve analysis may be conducted later using the method described in (M. Gritsevich & D. Koschny, 2011; A. Bouquet et al., 2014; E. Drolshagen et al., 2021a, 2021b), which allows for inference regarding the value of the shape change coefficient.

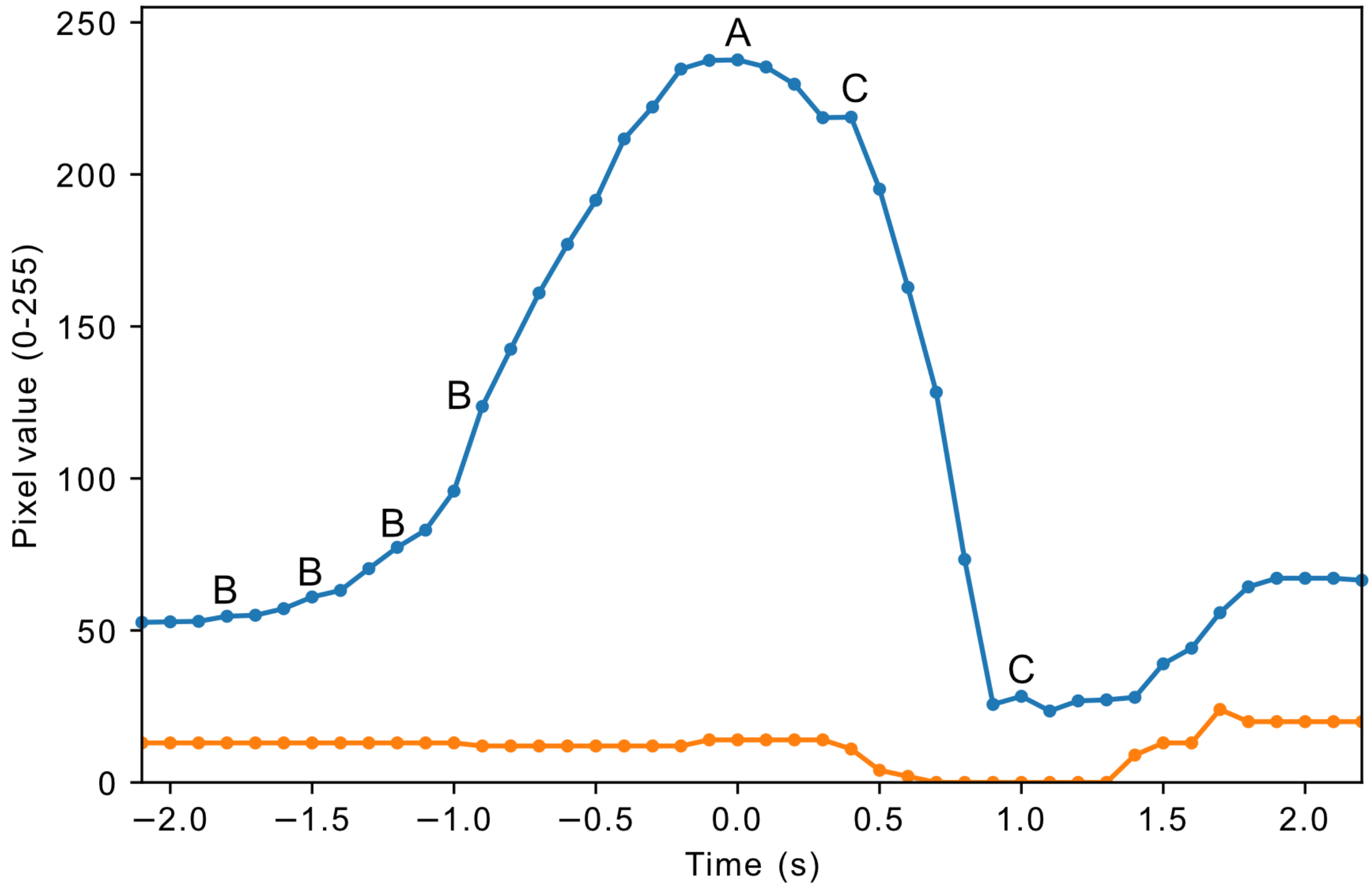

Figure 3. Secondary light curves from the Björkbacken video. The upper curve is average brightness in six different spots outside the house. The lower curve is the dark reference spot inside the house. The time axis shows deviation from the peak brightness A (0.0 s). Possible brightness variations due to rotation are marked as B and two possible flares are marked as C.

### 3.3 Timing the Audio

We analyzed the audio from the video using the Audacity 2.0.4 program[10]. The total duration of the video is approximately 1 minute and 23.1 seconds, with a margin of error of ±0.1 seconds due to the compressed mp4 format. Different audio-video programs provided varying lengths, with some indicating the video is 1 minute and 23.2 seconds long, while the AAC format audio itself is only 1 minute and 5 seconds. Notably, the video freezes at 1 minute and 13 seconds, which complicates precise event timing analysis.

In the video, at least eight distinct sonic booms were clearly audible (see Table 6, Fig. 4). Additionally, there were several periods of noise, likely caused by overlapping weak sonic booms, echoes, or background thundering sounds. When the audio was slowed down significantly, more than 50 boom-like sounds could be distinguished. These could be echoes of louder sonic booms, sounds from smaller fragments, or artefacts of the recording process. We assume that they are mostly echoes by pressure waves of main sonic booms.

Table 6. Timing of audio events recorded by a private security camera in Björkbacken. Time is set to zero at the peak brightening of the landscape, which was used to synchronize sound recordings with the trajectory. Asterisk (*) indicates audio events that could originate from three different masses.

| Time [mm:ss.ss] | Event descriptions |
| --- | --- |
| -00:05.34 | Start of the video. |
| -00:01.98 | First clear indication of light from the fireball. |
| 00:00.00 | Maximum brightness. Corresponds with NASA’s 22.3 km altitude. |
| (00:00.40) | Auto exposure of the camera kicks in. |

[10] https://sourceforge.net/projects/audacity/

| 00:01.00 | Still, some light from the fireball. |
|---|---|
| 00:01.12 | Dark outside again. |
| (00:09.00) | Exposure of the CCTV camera back to normal. |
| 00:26.97 | “The thud” sound. |
| 00:29.32 * | The 1st and loudest sonic boom. |
| - | A very loud rumbling sound. Including overlapping weak booms. |
| 00:32.10 * | The 2nd loud sonic boom. |
| - | More rumbling sound. |
| 00:33.64 | The 3rd clear but weak sonic boom. |
| 00:34.34 * | The 4th loud sonic boom. |
| 00:35.49 | The 5th sonic boom. |
| - | Loud rumbling sound. |
| 00:38.26 | The 6th sonic boom. |
| 00:39.13 | The 7th sonic boom. |
| 00:40.79 | The 8th clear but weak sonic boom. |
| - | Thunder-like noise continues. |
| 01:17.66 | End of the video. |

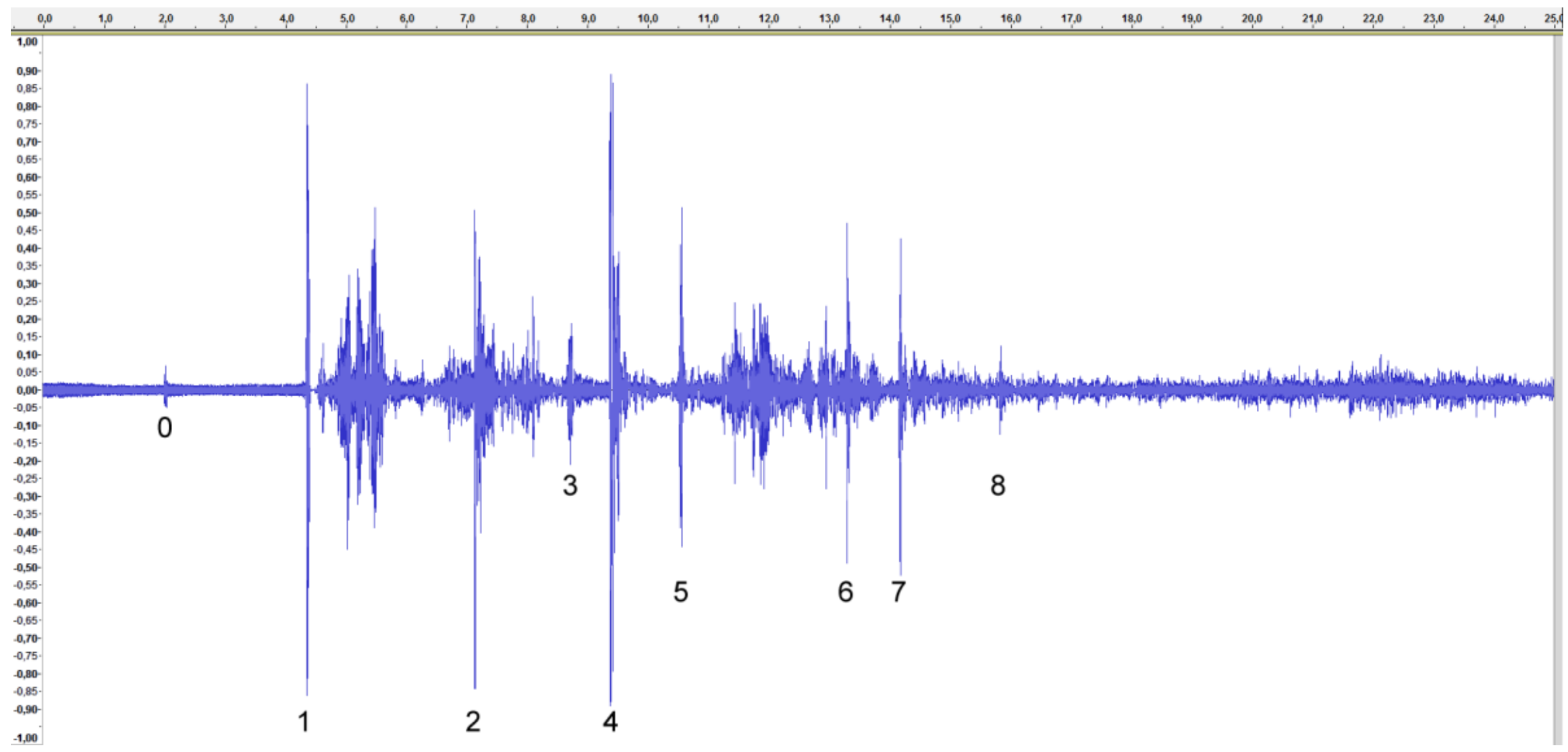

Figure 4. A graph of sound recorded by a private surveillance camera located in Björkbacken. Signal levels have been amplified. Sonic booms are numbered as in Table 6. Time 0.0 seconds at timeline is 25 seconds after the peak brightness at time 0.0 in Figure 3.

### 3.4 The Thud Sound

One notable sound occurs 2.35 seconds before the first sonic boom, marked as “0” in Figure 4. This sound resembles a thud. Possible explanations include:

1. It may be unrelated to the fireball and could originate from within the house.

2. It could be a seismic wave resulting from the meteorite impacting the ground.

3. It might be a seismic wave caused by the initial and strongest pressure wave reaching the ground.

Typically, such unusual sounds are not directly related to observed fireball events. However, in this instance, we consider it may be linked to the meteorite fall (explanation 3 above). The thud can be explained by a seismic wave caused when a strong pressure wave hits the ground. The speed of sound in bedrock is much higher (5950 m $s^{-1}$ in granite) than in the air, because of which the seismic wave arrives to the camera before the airborne pressure wave.

## 4 Methods and Results

### 4.1 Strewn Field

The initial trajectory of the fireball was reconstructed the following days after the event by Esko Lyytinen, primarily utilizing observations from Finland and Norway, employing the fb_entry program adjusted for real atmospheric conditions (E. Lyytinen & M. Gritsevich, 2013, 2016ab). Panu Lahtinen from the Finnish Meteorological Institute provided atmospheric data corresponding to the fall site for November 7, 2020, at 21Z (coordinates 60°N, 17°E, see Fig. 5) to aid in these computations. To further analyze the event in detail, including accounting for fragmentation and simulating the forecasted strewn field, we employed well-established DFMC model (J. Moilanen et al., 2021; M. Gritsevich et al., 2024). Atmospheric conditions over the impact area at the time of the meteorite fall are shown in graphs in Figure 5.

The DFMC model estimates the meteorite strewn field using a Monte Carlo approach in which multiple physical parameters are varied simultaneously within realistic limits. These include the initial conditions of the fragments, aerodynamic properties, and fragmentation parameters. In addition, stochastic variations in wind speed and wind direction are applied independently to each simulated fragment during dark flight. This reflects the fact that available atmospheric soundings represent spatially and temporally averaged measurements rather than the exact atmospheric conditions experienced by individual meteorites. Consequently, each simulated fragment encounters a different atmospheric realization during its descent. The resulting spread of the simulated landing positions therefore represents the combined uncertainty arising from observational errors, atmospheric variability, and model assumptions. Rather than a single propagated analytical uncertainty, the Monte Carlo ensemble itself provides the uncertainty estimate for the predicted strewn field.

Parameters used for DFMC simulations are provided in Table 7. The DFMC simulations were initially calculated from a starting altitude of 22.3 km corresponding to the peak brightness, which was later adjusted to 43.89 km (J. Moilanen & M. Gritsevich 2021) to better account for possible fragmentation events. This choice of altitude was constrained by the available atmospheric data, extending up to 44.6 km, with the aim of ensuring the strewn field prediction covered the potential extent of the fragment distribution. Because starting point for simulations is lower in the atmosphere than the beginning of the luminous flight, the initial velocity for DFMC simulation is slower than initial velocity for earth encounter.

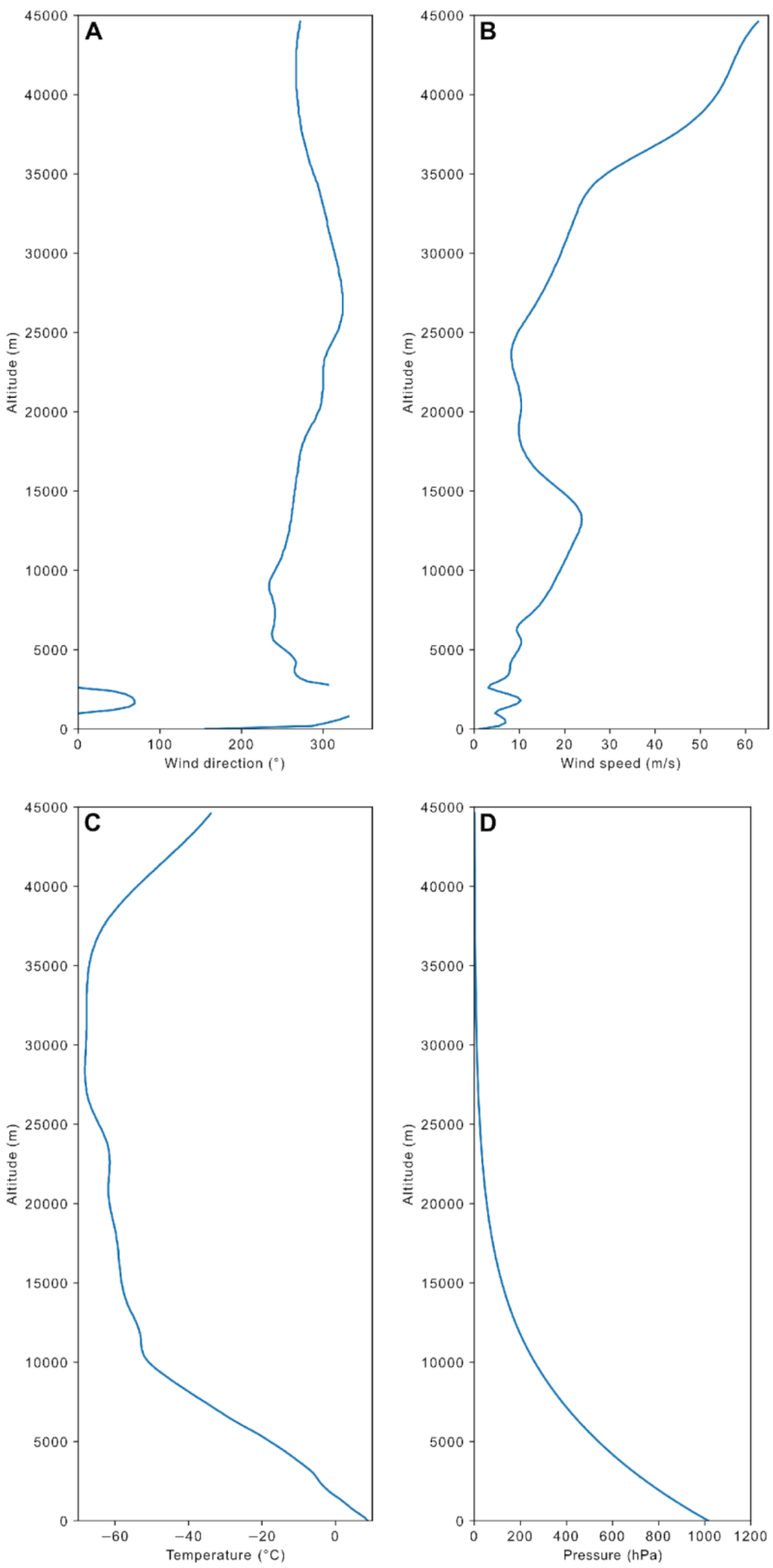

Figure 5: Atmospheric conditions over the impact area at the time of the meteorite fall, based on meteorological data provided by Panu Lahtinen (Finnish Meteorological Institute). Panels show

(A) wind direction, (B) wind speed, (C) air temperature, and (D) atmospheric pressure as a function of altitude.

In this study we utilized the most recent trajectory solution (I. Kyrylenko et al., 2023), retrieved using the FireOwl data-processing software of the Finnish Fireball Network (J. Visuri & M. Gritsevich, 2021; E. Peña-Asensio et al. 2024; M. Gritsevich et al. 2024). The solution is close to identical to the earlier trajectory derivation made by Esko Lyytinen. Additionally, trajectory parameters from Eric Stempels (personal communication, 2021) were initially considered. However, the assumption that the 13.8 kg meteorite fell precisely along the center line of the strewn field, which was incorporated as an input condition in this later solution, was subsequently discarded. This decision arose from concerns that this approach might not accurately reflect the actual trajectory solution.

Table 7. The DFMC input parameters retrieved using the FireOwl data-processing software for the iron meteorite fall in Sweden on November 7, 2020.

| Symbol | Parameter | Mean value | Error margins | Remarks |
|---|---|---|---|---|
| $\lambda_0$ | longitude | 16.68715°E | N/A | WGS84 (error margins included in $e_0$) |
| $\varphi_0$ | latitude | 59.78321°N | N/A | WGS84 (error margins included in $e_0$) |
| $h_0$ | altitude | 43 890 m | N/A | Error margins included in $e_0$. |
| $e_0$ | spatial error at the start | 0 m | ±300 m | Default error margins. |
| $\delta_0$ | direction of trajectory | 61.877° | ±1.8° | 0° - 360° (0° = N, clockwise) |
| $\gamma_0$ | trajectory slope | 73.364° | ±1.8° | 0° - 90° (90° = vertical) |
| $V_0$ | velocity | 17 341 m $s^{-1}$ | ±300 m $s^{-1}$ | Default error margins. |
| $a_0$ | deceleration limit | 302 m $s^{-2}$ | $a \geq 0.99*8$ m $s^{-2}$ | Deceleration at the starting point. |
| $h_{eh}$ | end height of fireball | 11 280 m | - | Lowest observed altitude. |
| $\rho_m$ | density of meteoroid | 7500 kg $m^{-3}$ | ±500 kg $m^{-3}$ | Default for iron meteorites. |
| $\sigma$ | ablation coefficient | 0.0018 $s^2$ $km^{-2}$ | - | Our default. |
| $t_a$ | ablation temperature | 1 550 °C | - | Our value for iron. |

Figure 6 (left) shows the result of a single DFMC simulation, performed using the method described in J. Moilanen et al. (2021). This simulation produced nine fragments with terminal masses ranging from 0.10 to 13.8 kg. The total cumulative mass on the ground in the simulation was 31.3 kg. The mass of the meteoroid at the starting point of the simulation, derived from the observed deceleration, was 47.21 kg, resulting in a surviving nominal mass of 13.8 kg. This nominal mass (square symbol in Fig. 6) is located approximately 440 m from the actual impact point of the recovered iron meteorite (star symbol). In this nominal case, the main mass ablates, but does not undergo further fragmentation, and the wind profile is taken directly from atmospheric data without Monte Carlo variations (J. Moilanen et al. 2021, M. Gritsevich et al. 2024).

Figure 6 (right) presents the combined results of ten separate DFMC simulations, excluding the single simulation shown on the left. Together, these ten runs produced 114 fragments with masses ranging from 0.01 to 22.0 kg. On average, the total mass that reached the ground is 34.6 kg per simulation.

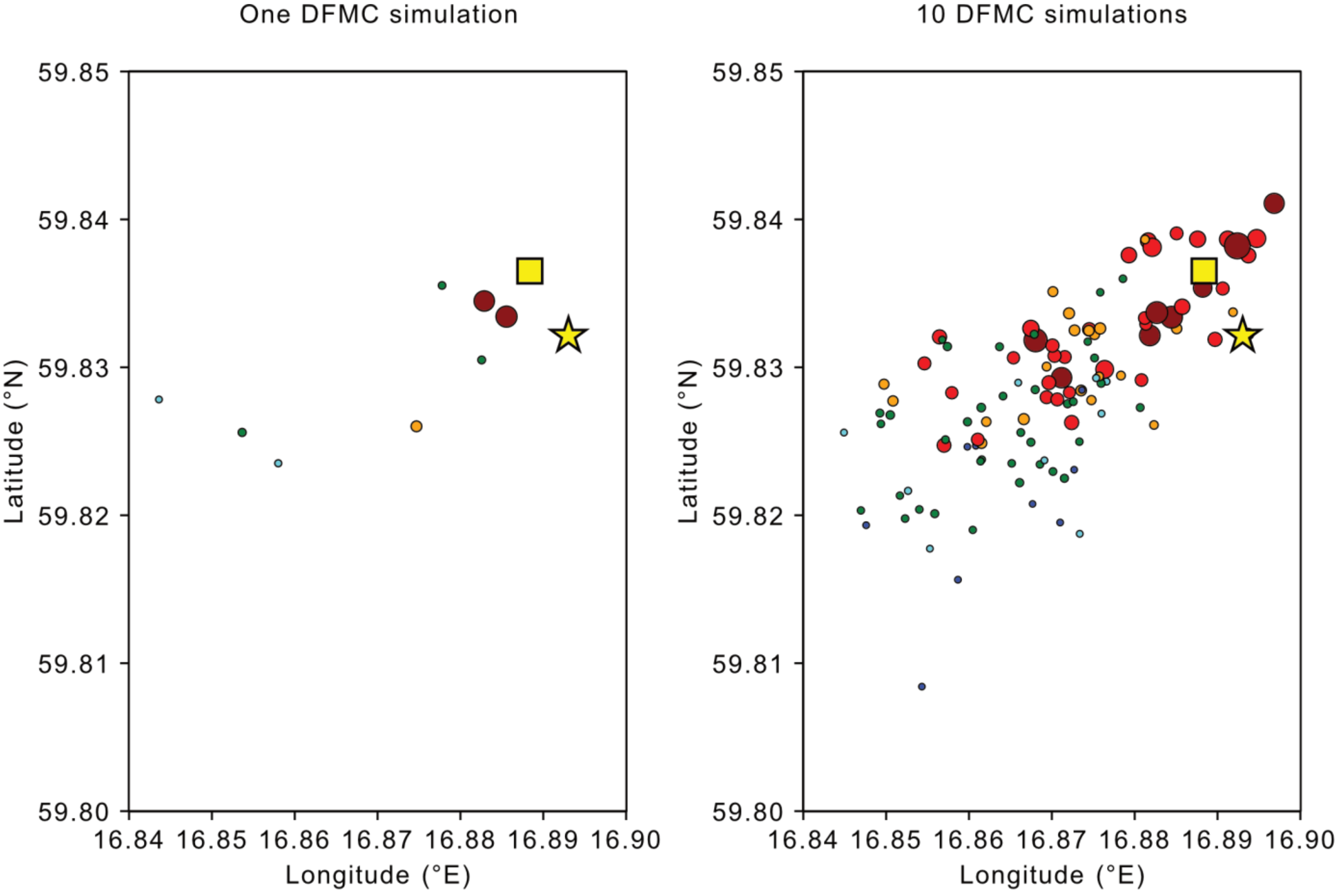

Figure 6. Left: A single DFMC simulation initiated from an altitude of 43.89 km. Right: Ten combined DFMC simulations, also starting from 43.89 km, to encompass the full extent of the potential strewn field. The color coding of the simulated fragments is as follows: blue <100 g, cyan 100–300 g, green 0.3–1.0 kg, orange 1–3 kg, red 3–10 kg, and dark red >10 kg. The surviving nominal mass (square symbol) is 13.8 kg. This nominal mass is located approximately 440 m from the actual impact point of the recovered iron meteorite (star symbol).

### 4.2 Effect of the Fragment's Shape

The recovered 13.8 kg iron meteorite fragment exhibits an exceptional and distinctive morphology. The specimen is a compact, irregularly shaped iron meteorite with a distinctly asymmetric, lopsided mass distribution (Fig. 7). Its overall form resembles a distorted wedge or knobby ellipsoid, with one side more strongly convex and the opposite side flatter and broken by shallow concavities. The reported meteorite's longest axis is approximately 30 cm. In plain view, it appears broad and rounded at one end and narrower and more tapering at the opposite end, giving the sample a directional, flow-like geometry. The width and thickness vary markedly along its length, producing a clearly non-uniform, unbalanced outline rather than a centrally symmetric body.

The surface is dominated by well-developed regmaglypts. These are shallow, irregular, and overlapping, ranging from small pits to broad, smoothly curved hollows several centimeters across. A continuous fusion crust covers most of the surface. One flank exhibits a prominent mechanical break and associated crack system, appearing as a jagged, irregular fracture that cuts across the otherwise smoothly ablated exterior. This break does not follow the aerodynamic sculpting of the rest of the meteorite and is therefore interpreted as secondary damage, consistent with impact against the granite boulder described in the find context. In this fractured zone, regmaglypts are locally truncated and the fusion crust is interrupted, producing sharp angular edges that contrast with the otherwise rounded morphology.

Three additional minor fractures occur on the same side of the meteorite. These are open tensile fractures, suggesting that the impact force was applied from the opposite side during collision with the boulder.

In side view, the meteorite displays a low, thick profile, with one side forming a broad, gently domed surface, while the opposite side is more uneven. The overall volume is off-center toward the broader end. This aerofoil-like shape significantly reduces turbulence and drag by enabling air to flow more smoothly over its surface. Consequently, the shape does not align with the drag coefficient range typically assigned to our default fragments of less exotic shapes, which assume $C_d = 1.5 \pm 0.5$ (J. Moilanen et al. 2021). To test the effect of variation in shape we made simulations with $C_d$ values of 3 ±0.5 and 4 ±0.5. In our DFMC code $C_d$ value of 2 corresponds to the motion of a bowl-shaped object and $C_d = 5$ stands for a flat plate.

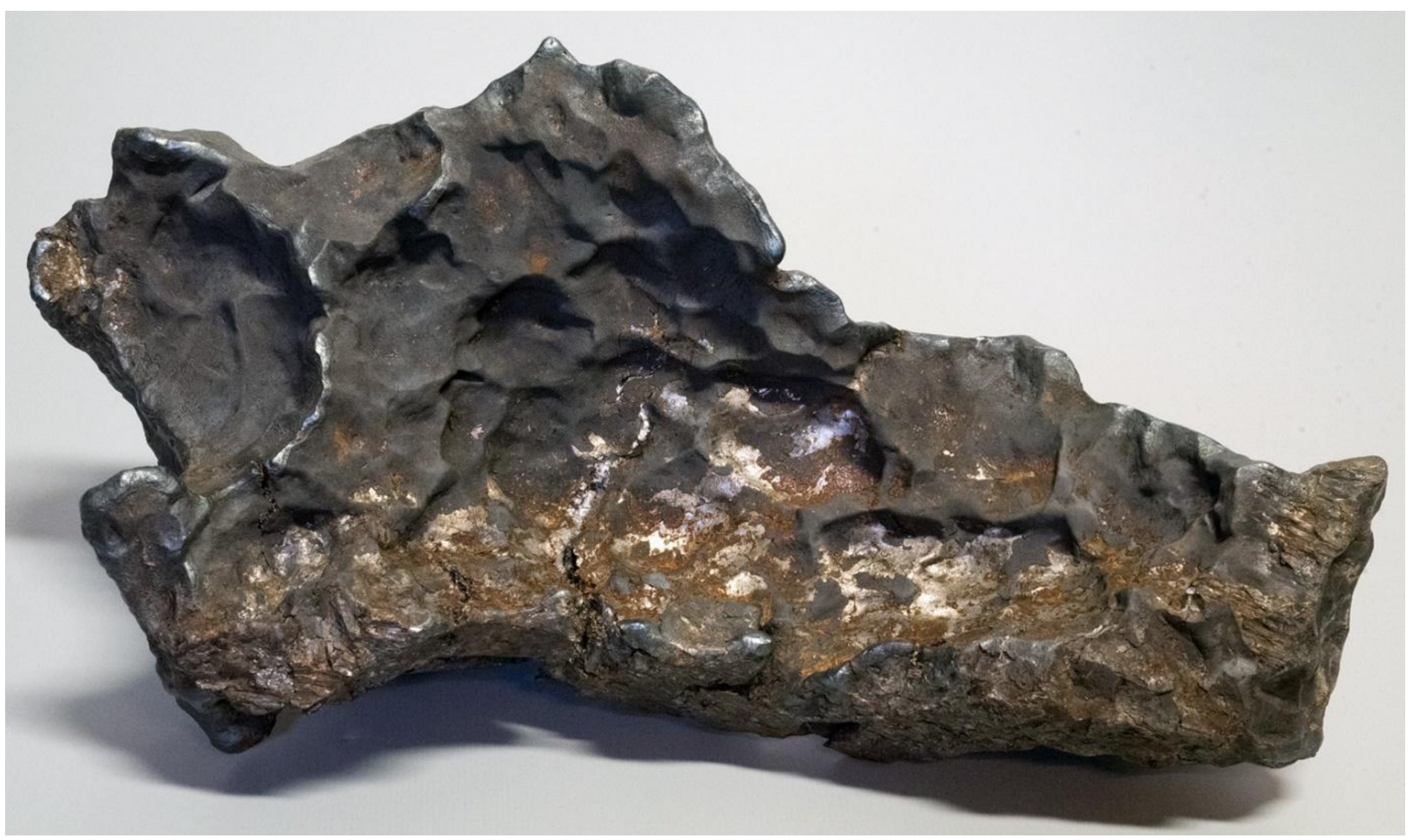

Figure 7a. Iron meteorite that fell in Sweden on 2020 November 7 is approximately 30 cm long and weighs 13.8 kg. The meteorite exhibits classic iron-meteorite ablation features, overprinted by localized impact fracture caused by its collision with terrestrial rock after atmospheric deceleration. Image published with permission of © Andreas Forsberg & Anders Zetterqvist.

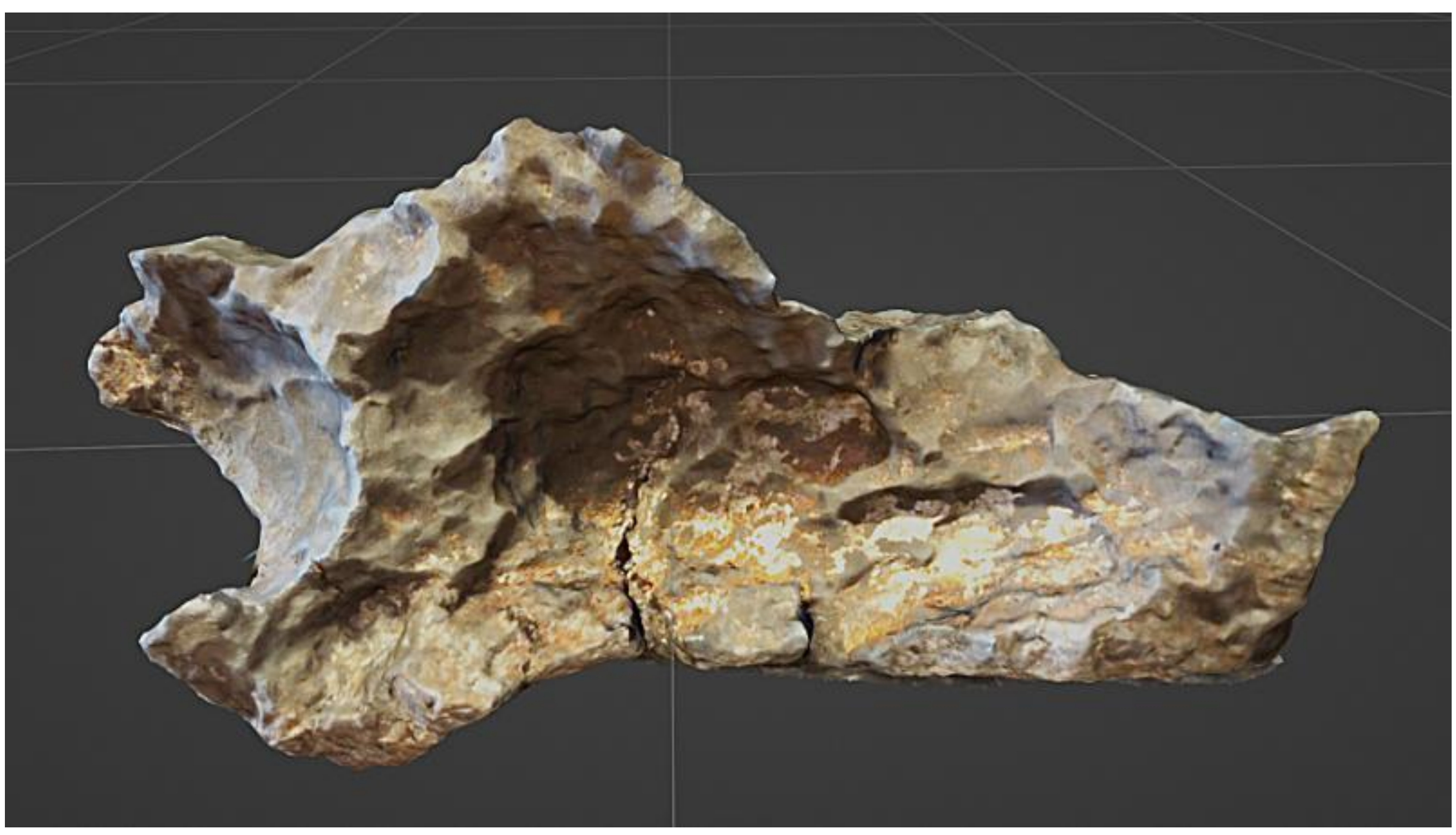

Figure 7b. View of the partial three-dimensional model of the meteorite reconstructed by the authors using photogrammetry from a video recorded by Clas Svahn. The 3D model is provided as Supplementary Material.

Simulations with larger $C_d$ values show that even large masses will hit the ground earlier than in the scenario with the default $C_d$ value. For instance, with a $C_d$ value of 4, the landing area for fragments weighing over 10 kg shifted nearly one kilometer south-westward and widened considerably. This shift occurred because the free-fall time increased significantly for similar-sized masses when compared to our default $C_d$ value.

The DFMC code does not consider possible drifting caused by asymmetrical shapes. When a fragment's center of mass differs significantly from its geometric center, noticeable horizontal drifting may occur. Additionally, such fragments might experience noticeable lifting or steering forces, introducing uncertainties in trajectory calculations. Further investigating these aerodynamic complexities may be advantageous for assessing the full impact of such factors on trajectory predictions.

The original purpose of the DFMC code was to generate high-probability distributions of fallen meteorites without the need to speculate about shape-dependent forces (J. Moilanen et al., 2021). Its aim is to produce guiding maps for recovery teams as soon as possible, prior to any meteorite recovery. Default error margins in DFMC parameters were adjusted based on known strewn field distributions of recovered fragments. Variations in fragment shapes are approximated by applying different drag coefficients for each fragment. Incorporating shape-dependent forces, such as the Magnus effect or aerodynamic lift, is challenging because the actual shapes of fragments are unknown before recovery. Adjusting the dark flight trajectory with data from recovered samples is therefore a part of post-event analysis, not addressed in our simulations.

Our DFMC simulations show that the ground trajectory deviates approximately 400 meters from the impact point, which is a rather accurate match considering the meteorite shape. Historical data from other meteorite falls indicate that even large fragments can land far from ground projection of their predicted trajectory or from the highest-probability line defined by earlier dark flight approaches. During the fall of the St. Michel chondrite in Finland on July 12, 1910 (L. H. Borgström, 1912), two pieces weighing 9.6 kg and 6.8 kg were found 2.3 km apart from each other, nearly perpendicular to estimated trajectory of the fireball. Similarly, after the Hessle meteorite fall in Sweden on January 1, 1869, several large fragments were discovered more than 2 km away from the central line of the main strewn field (A. E. Nordenskiöld 1870). Deviations in landing positions are influenced not only by the shapes and orientations of individual meteorites, but also by subtle differences in the atmospheric conditions each fragment encounters during atmospheric descent.

### 4.3 The Boulder

One remarkable peculiarity of the 13.8 kg iron meteorite impact on the ground is that it is inferred to have struck a large boulder (WGS84 coordinates: 59.83209° N, 16.89299° E ±10 m) before ending up in its final landing position. Small meteorite fragments were recovered in the vicinity of the boulder, indicating that the fragments broke off the meteorite upon impact (J. Moilanen & M. Gritsevich 2022). The resting place of the meteorite was in the SSW direction (at an azimuth of 200°) from the boulder (WGS84 coordinates: 59.83148° N, 16,89256° E ±10 m). The presence of small fragments around the impact site confirms that the meteorite sustained damage when it hit the boulder. Additionally, the 13.8 kg iron meteorite itself exhibits a clear crack in its middle area (Fig. 7).

Visible damage on top of the boulder corroborates the impact (Fig. 8). The measured distance between the boulder, the actual impact point, and the meteorite's terminal resting place (Fig. 10) is approximately 75 ±1 meters. Interestingly, the resting spot is about 3 meters higher (74 meters above sea level) than the impact point (Fig. 11). Despite the presence of numerous large and tall trees between these two points, no hit marks were found on the trees during onsite examination in April 2021, suggesting that the meteorite managed to miss them all during its flight.

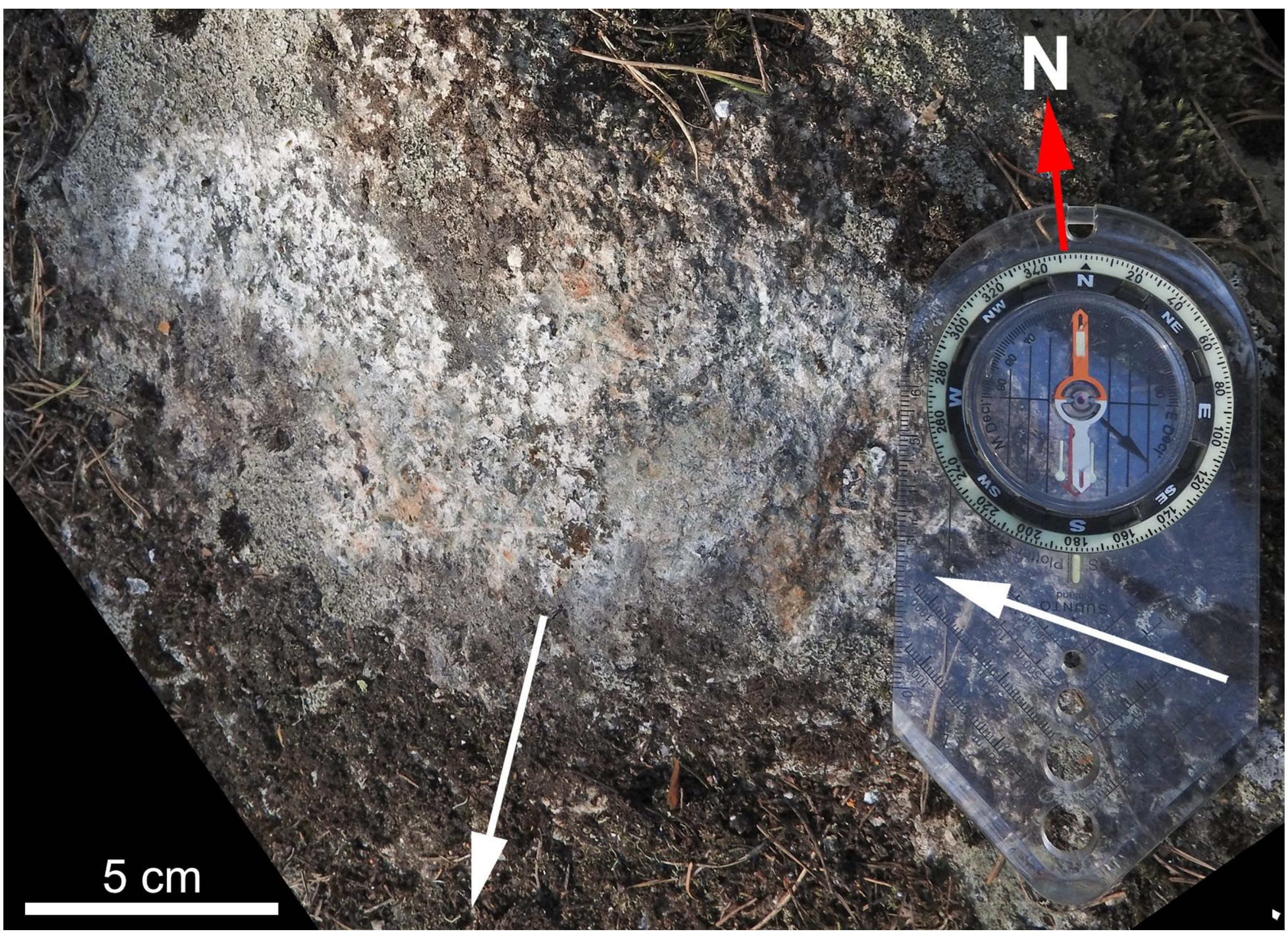

Figure 8. A light-colored scar is visible on the top of the boulder, photographed on 19 April 2021.

In Figure 8 the geographical north is indicated by a red arrow. The compass needle points to the magnetic north. This part of Sweden the magnetic declination is +6°, which has been used to draw the arrows in the correct directions. The white arrow from the right shows the impact direction (295°) as obtained from a DFMC simulation, though this direction is not relevant given the nearly vertical trajectory. Between white arrows is an approximately 5 cm wide deeper impact dent on the boulder, where striations caused by the impacting meteorite are visible. The white arrow towards down left points to the direction (200°) where the meteorite was recovered.

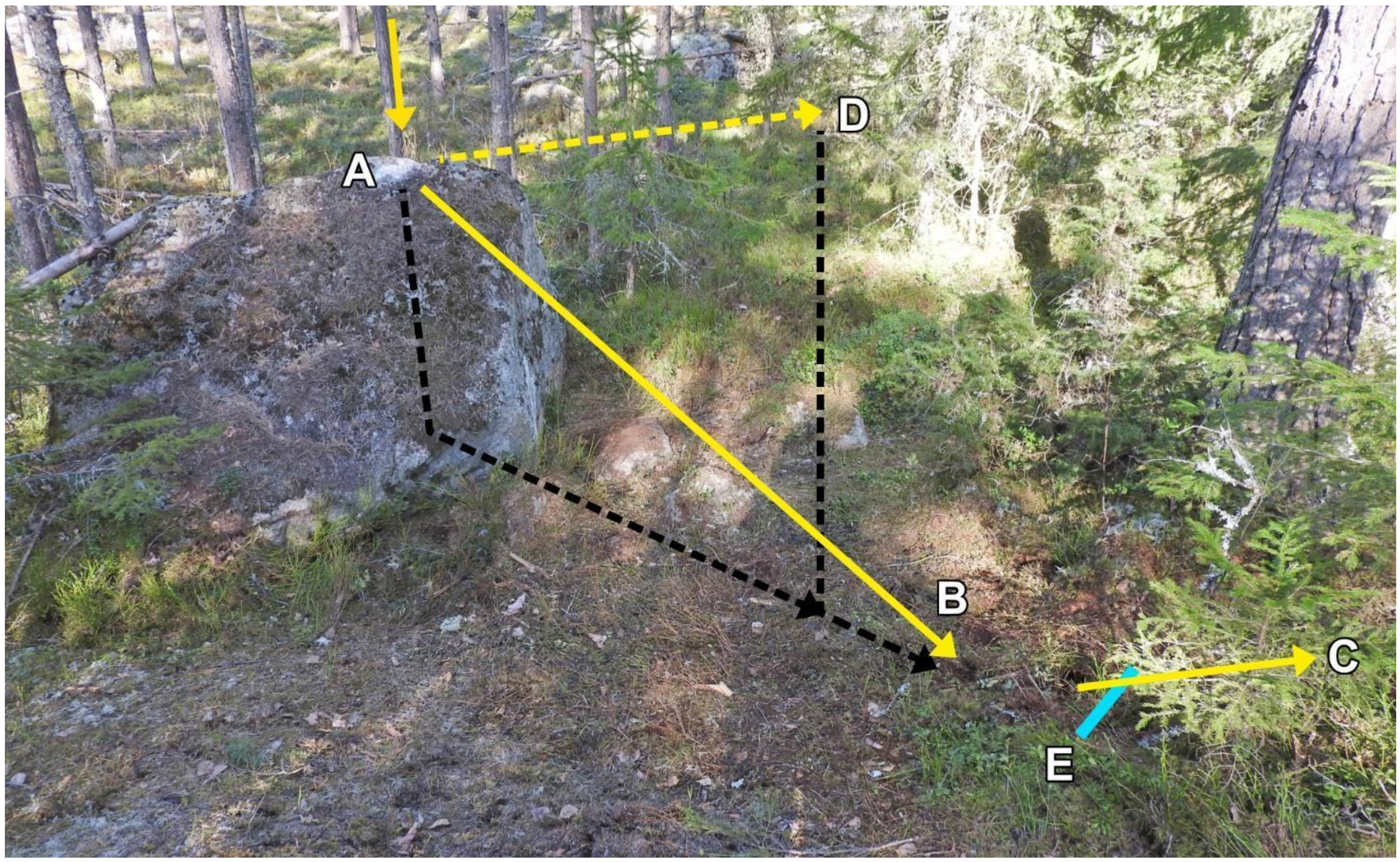

Figure 9. The boulder and two possible trajectories of ricochet, A-B-C and A-D, originating from the impact scar at A. Trajectory A–B–C represents a path including a secondary ricochet off the ground between points B and C, while trajectory A–D represents a direct airborne path to the in-situ location. The light blue line (E) indicates the pine root. Black lines show the ground projections of trajectories A–B and A–D.

In Figure 9 arrow D shows a ricochet trajectory upward from the scar and arrow B is the ricochet trajectory towards the ground. The arrow C is direction after the ricochet from the ground.

Arrows B, C and D are pointing in the same azimuth (220°) from the boulder. The pine root, seen in the upper right corner, lies approximately halfway between points B and C, marked by the blue line (E). The distance from the impact scar on the boulder to the small pine tree is about 3 meters.

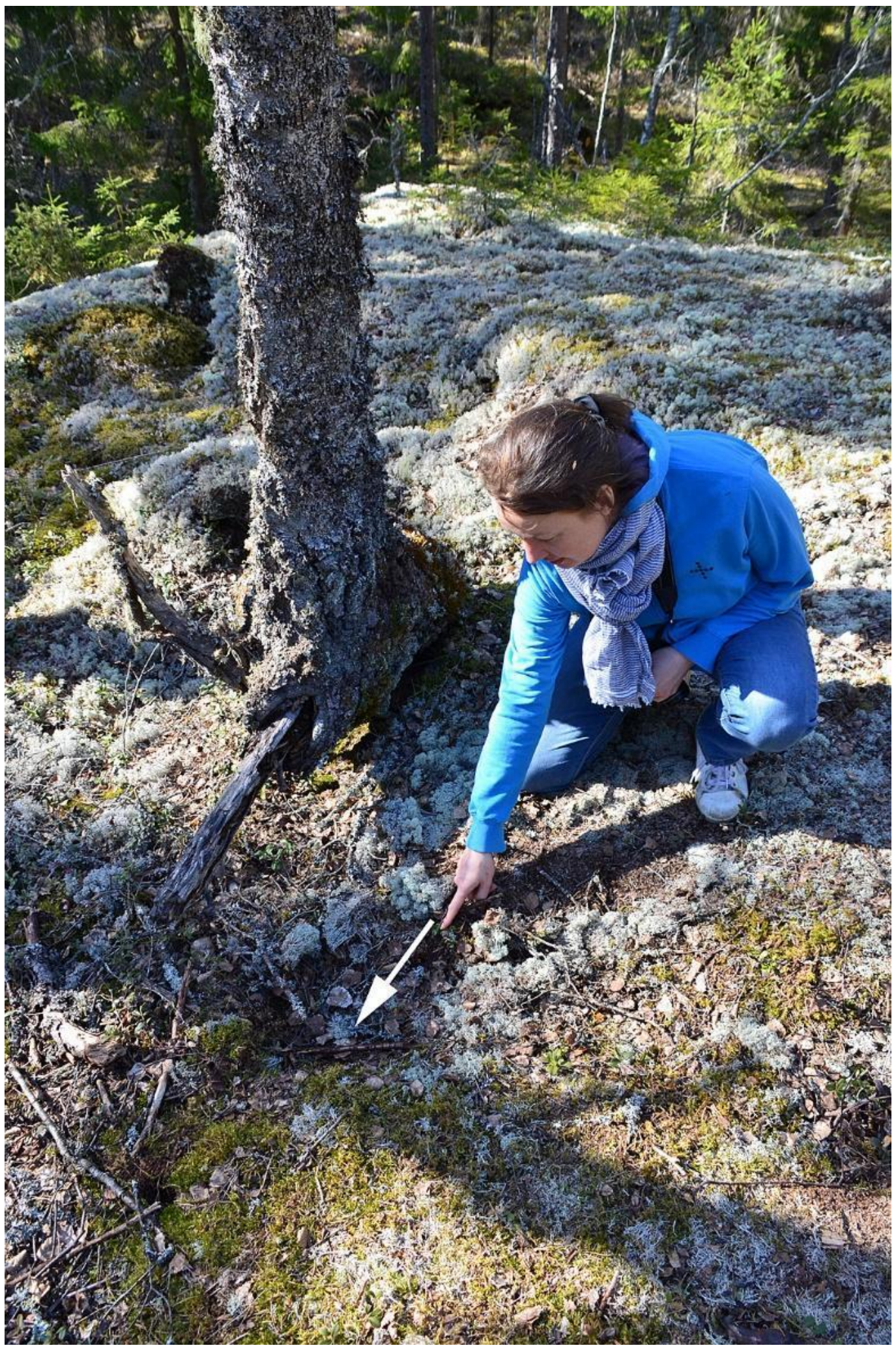

Figure 10. The white arrow points to the location where the iron meteorite was discovered, indicating both its resting place and the approximate direction from which it traveled after striking the boulder 75 meters away.

The point where the meteorite initially struck was nearly the top of the granite boulder, approximately 1.3 meters above the ground (about 71.8 m above sea level). The erratic boulder measures approximately 2.6 meters by 1.8 meters in horizontal dimensions and stands about 1.5 meters tall. It has a slightly skewed triangular prism shape. The meteorite impacted near the top wedge of this prism, on a side surface with a slope of approximately 47° from the vertical. The impact scar on the boulder measures approximately 22 cm in length and 9 cm in width, located on the southwest-facing slope. The scar is not flat but spans over two surfaces that form an angle of about 143° between them, as depicted from a 3D model.

The largest surface of the boulder struck by the meteorite was inclined by approximately 47° from the vertical. Assuming a nearly vertical impact trajectory, the meteorite would have rebounded from the boulder at an angle of approximately 4° above the horizontal.

After rebounding, the meteorite came to rest on a small bedrock outcrop adjacent to a birch tree, as documented in the in-situ photographs (Fig. 10). It became lodged beneath a narrow birch root without causing any visible damage to the tree. This is consistent with the very shallow rebound trajectory resulting from the impact with the inclined boulder.

We made simulations using the DFMC code to determine if the 13.8 kg iron meteorite could have created the observed hit mark on the boulder. Various simulations with different ricochet angles for the iron meteorite were run. In these simulations iron fragments weighting 12 – 16 kg impacted the ground at velocities 90 – 110 m $s^{-1}$, depending on their shape. Variations in fragments' shape are accounted for by variation in the drag coefficient (J. Moilanen et al., 2021; M. Gritsevich et al., 2024). We selected 100 m $s^{-1}$ as the mean impact velocity for the meteorite, although the shape of the found meteorite suggests a possibly lower impact velocity. In these simulations, the average impact angle of the fragments with the ground was 1.88° ±0.2° from the vertical, while the mean flight direction was 295.2° ±20°. It is important to note that in the DFMC simulation, other forces related to shape and motion, such as the Magnus effect, are ignored (J. Moilanen et al., 2021; M. Gritsevich et al., 2024).

It is not certain if the heavy boulder lies on a shallow moraine bed or directly on local bedrock. Although bedrock is visible just a few meters from the boulder, the area immediately around the boulder consists of softer soil and smaller rocks. We assume the boulder is firmly anchored to the ground and did not move due to the meteorite impact. In an inelastic collision, kinetic energy is not conserved, although momentum is. Some of the kinetic energy is transformed into heat and sound. The coefficient of restitution (COR, $e$) for solid stainless-steel spheres ranges from 0.63 to 0.93, and it is less for cast iron. Experiments for granite have yielded $e = 0.83 \pm 0.06$ (D. D. Durba et al. 2011). Initially, as a boundary condition, we applied ricochet velocity (Vr) of 100 m $s^{-1}$ corresponding to $e = 1$, which is naturally unrealistic. Using Vr = 100 m $s^{-1}$ as a ricochet velocity, we found that with a ricochet angle $\delta = 4°$ from the horizon plane, a 13.8 kg iron meteorite can indeed fly 75 meters and land approximately 3 meters higher. This suggests that the iron meteorite found 75 meters from the boulder and the small fragments found nearby were probably from the same meteorite. However, the re-impact velocity at the final resting place would have been as high as 93 m $s^{-1}$ (= 334.8 km $h^{-1}$). At such a speed, the iron meteorite likely would not have come to rest where it was found. The same reasoning applies to the scenario in which the meteorite landed directly at its recovery site: a high-impact velocity would have produced noticeable damage to the ground. Additionally, achieving the low angle required to pass beneath the thin birch root would have been virtually impossible if the meteorite had arrived there directly from free fall.

Calculating the exact deflection or ricochet angle is challenging due to numerous variables in the collision, including the shape, motion, and mass center of the meteorite, as well as the shape and strength of the boulder. Therefore, other ricochet angles are also plausible. According to a 3D model, created using photogrammetry, the hit mark on the boulder is not on a steep slope, making a higher flight path possible after deflection. For example, a ricochet velocity of 50 m $s^{-1}$ with $\delta = 11°$ would allow a 14 kg iron meteorite to land at 75 meters distance on a 74 meters elevation with a 46 m $s^{-1}$ impact velocity (Figure 14). Other solutions are likewise possible, and Figure 11 also shows a trajectory for a ricochet velocity of 36.5 m $s^{-1}$ with $\delta = 20°$. This is a more plausible flight path from the impact scar on the boulder, as the slope in the direction of the resting place of the meteorite is approximately 35° from the vertical.

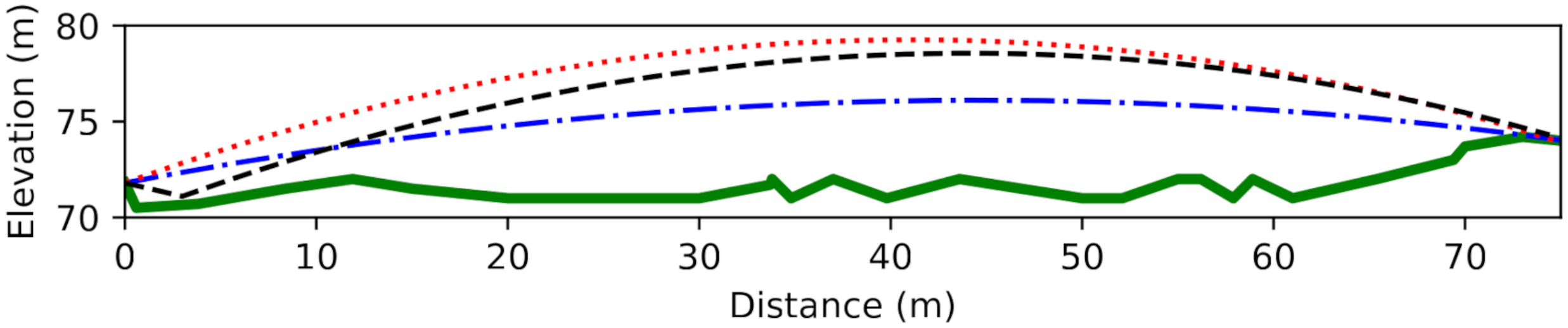

Figure 11. Elevation profile approximation of the ground (green solid line) between the initial impact point on the top of the boulder (distance 0 m) and the final resting place of the 13.8 kg iron meteorite (distance 75 m). Elevation data are extracted from an online map[11] by Lantmäteriet.

In Figure 11 the black dashed line shows the trajectory of the meteorite from the boulder via ground contact with Vr = 36.5 m $s^{-1}$ and δ = 20°, corresponding to trajectory A-B-C in Figure 9. The blue dash-dot line represents the flight path of the meteorite from the boulder when the ricochet velocity (Vr) is 50 m $s^{-1}$ and the ricochet angle δ = 11°. The red dotted line is the flight path when Vr = 36.5 m $s^{-1}$ and δ = 20°. Both approximately match trajectory A-D in Figure 9. Fig. 12 presents the results based on ballistic trajectory equations (Eq. 1) without considering air drag. For Figure 12 values $x$ = 75 meters, $y$ = 3 meters and $v$ from 0 to 160 m $s^{-1}$ were applied.

Eq. 1 $$\theta = arctan\left(\frac{v^2 \pm \sqrt{v^4 - g(gx^2 + 2yv^2)}}{gx}\right)$$

According to such a level of estimates, all ricochet angles are plausible. The ricochet velocity needs to be up from Vr = 28.8 m $s^{-1}$ to reach the final resting place of the meteorite when ricochet angle δ = 45°. By assuming the likely ricochet angle to be in the range 5° – 25°, the probable ricochet velocity is in the range 31.5 – 65.5 m $s^{-1}$. Air drag makes much lower angles unrealistic and initial velocities slightly higher. It also makes the trajectory steeper at the end, therefore high ricochet angles are unrealistic, as also explained above.

[11] https://minkarta.lantmateriet.se/

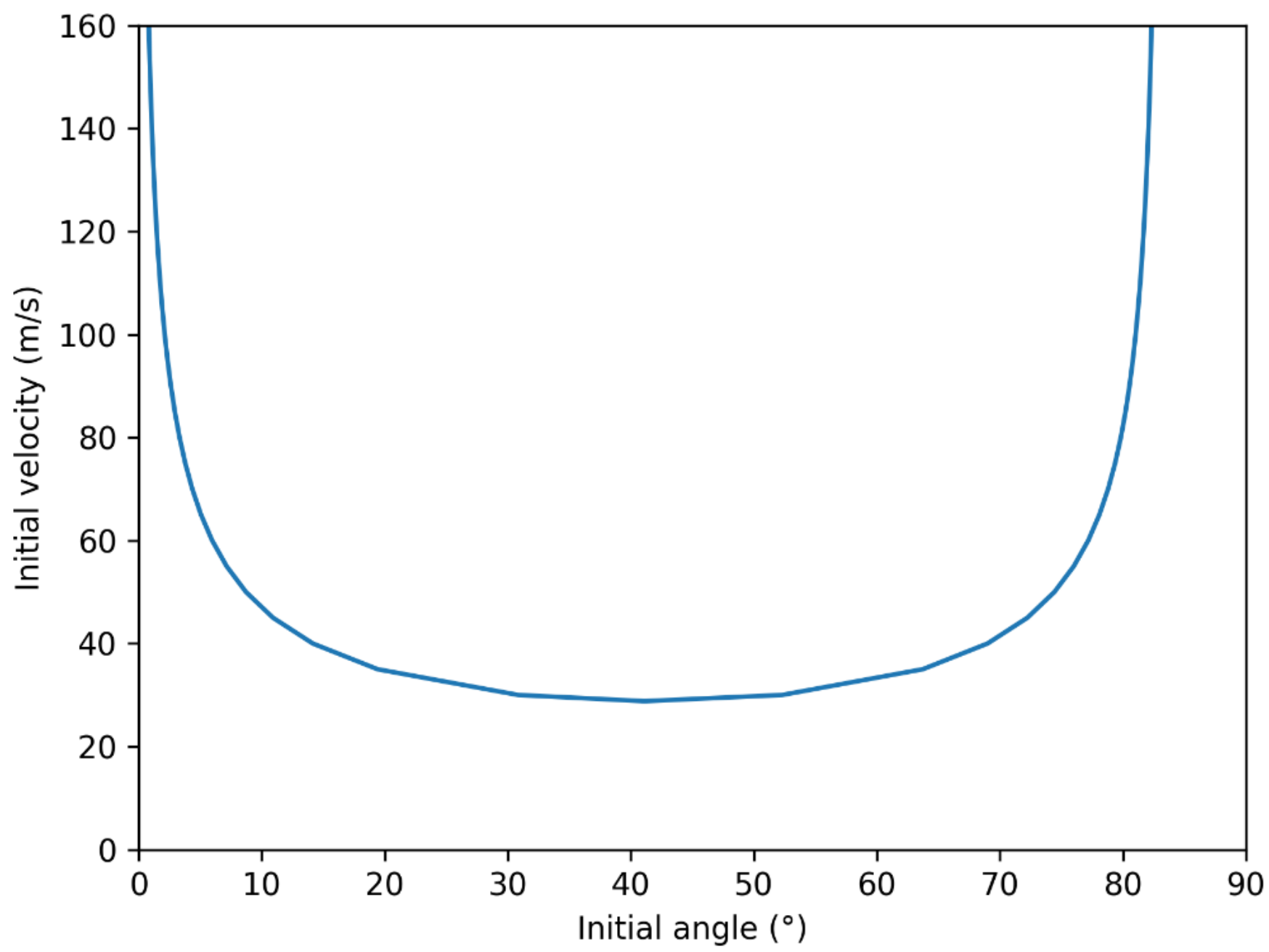

Figure 12. The initial ricochet angle versus the initial ricochet velocity required for a meteorite to land 75 meters from the boulder and three meters higher. The curve is calculated using basic ballistic equations (Eq. 1) without accounting for air drag during the ricochet phase.

### 4.4 Did the Meteorite Strike the Ground Near the Boulder Too?

It has been suggested by Swedish researchers that the meteorite slid off the boulder upon impact and subsequently bounced off the ground, traveling all the way for approximately 72 meters (M. Testorf 2021a). There is 3 meters between the initial impact point on the boulder and the first suggested trace on the ground, which appears to point in the same direction as the subsequent ricochet flight path. There are challenges with this interpretation.

Wild boars (*Sus scrofa*) are known to dig in the soil, creating similar ditches that can be seen all over the area. Complicating the analysis, there was another similar trace in the soil only a meter away, both of which had the same direction. Furthermore, a wide and thick intact root of a nearby pine tree cuts across the ditch thought to be caused by the meteorite (blue line (E) in Figure 9). It is highly improbable that a meteorite could pass by the root, leave it intact, and still retain enough momentum to travel over 70 meters. However, we cannot rule out the possibility that the other ditch, or part of the ditch intersecting the root, was created by meteorite hunters rather than by the meteorite itself.

To quantitatively explain this scenario, we performed DFMC simulations to determine the ricochet velocity needed for an iron meteorite to jump over 70 meters from the ground to a location more than 3 meters higher, given that the ditch on the ground is 0.7 to 0.8 meters lower than the hit mark on the boulder. These simulations aim to provide a plausible explanation for the ricochet trajectory and resting place of the meteorite.

We conducted a DFMC simulation to assess whether a jump from the ground, as shown in Figure 14, could be feasible. For this, we used the same parameters as in the second ricochet flight path: an initial velocity Vr = 36.5 m $s^{-1}$ and a launch angle of $\delta = 20°$ from the horizon. This already gave a very good match. The lowest velocity needed the meteorite to reach the final resting place, using a simple ballistic equation, is approximately 28.8 m $s^{-1}$ at a launch angle of about 45°. If the impact velocity with the boulder was 100 m $s^{-1}$, a post-impact velocity of about 35 m $s^{-1}$, potentially after carving a ditch in the ground, seems plausible. However, this scenario remains untested. A remaining question is how the meteorite passed through the thick pine root without shattering it. The most probable explanation is that the root yielded just enough to allow the meteorite to pass, leaving it largely intact. Arrow C in Figure 9 marks the location of this interaction.

Conversely, if the fragments were scattered around the boulder, not in the ditch, as reported by M. Testorf (2021a), there is less reason to associate the ditch with the meteorite's trajectory. The notion of the meteorite striking the ground near the boulder and continuing to its final location may seem far-fetched. Nevertheless, we acknowledge that such an event, while improbable, is not impossible. Meteorite skipping over long distances has been observed; in 1902, the Marjalahti pallasite struck bedrock, and its largest fragment landed in water about 50 meters away after bouncing twice (L. H. Borgström, 1903). Such skipping behavior is probably more likely with iron and stony-iron meteorites than with more fragile stony meteorites.

## 4.5 Matching the Trajectory with Audio

An additional critical dataset comes from a seismograph recording taken at five seismometers (BACU in Vittinge, FIBU in Fiby, OSTU in Horsskog, FLYU in Österbybruk and ESKU in Eskilstuna) located at 15 – 70 km distance from the fall site (Fig. 13). The coordinates of the sound source can be inferred by analysing the source distances from the seismometers using the coordinates of the Swedish seismograph network (SNSN 2020)[12]. The recordings do not reflect seismic ground waves, but pressure waves propagating through the air.

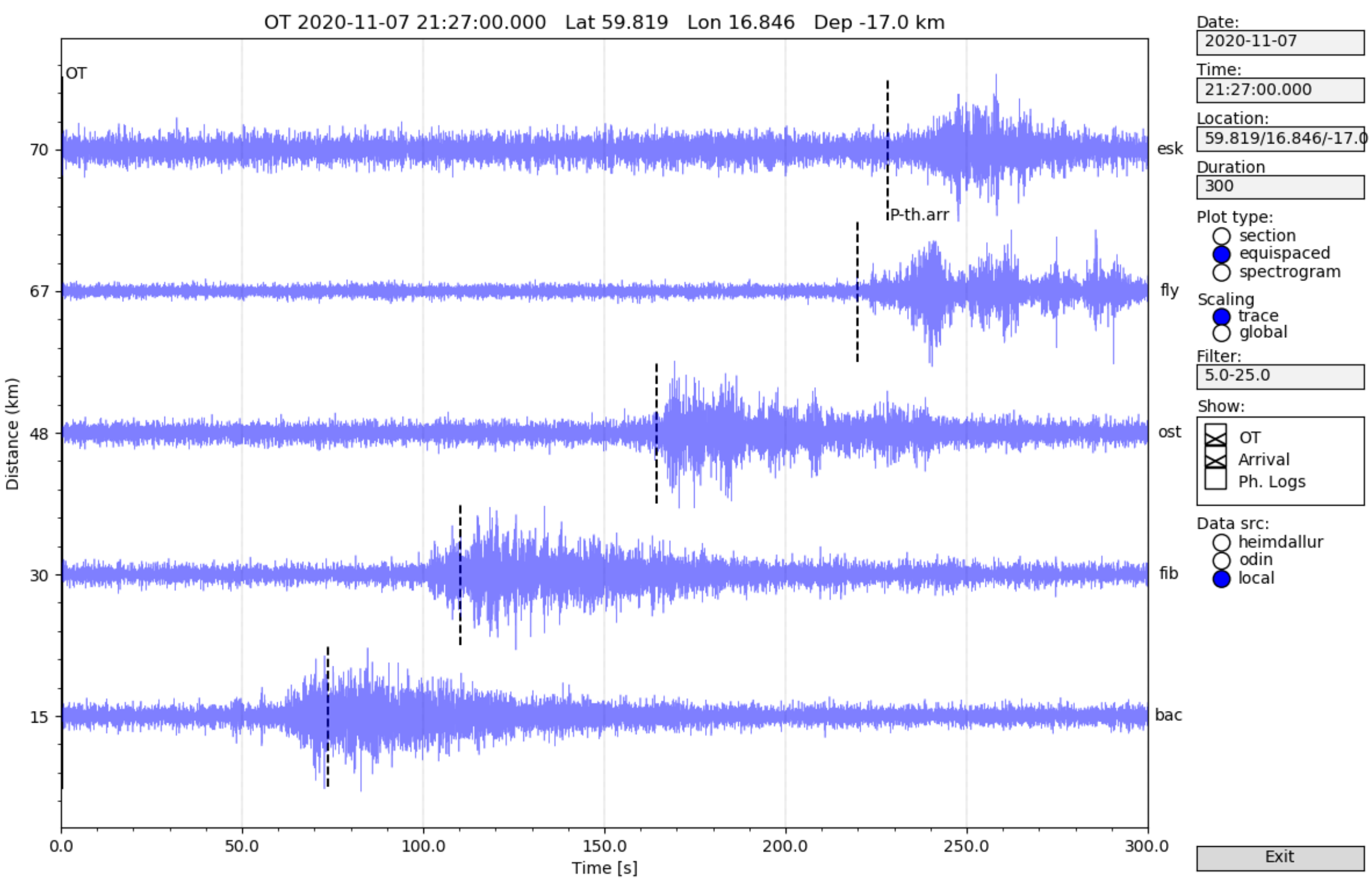

Figure 13. Seismic data from five different seismic station in Sweden (SNSN 2020).

[12] https://fdsn.org/networks/detail/UP/

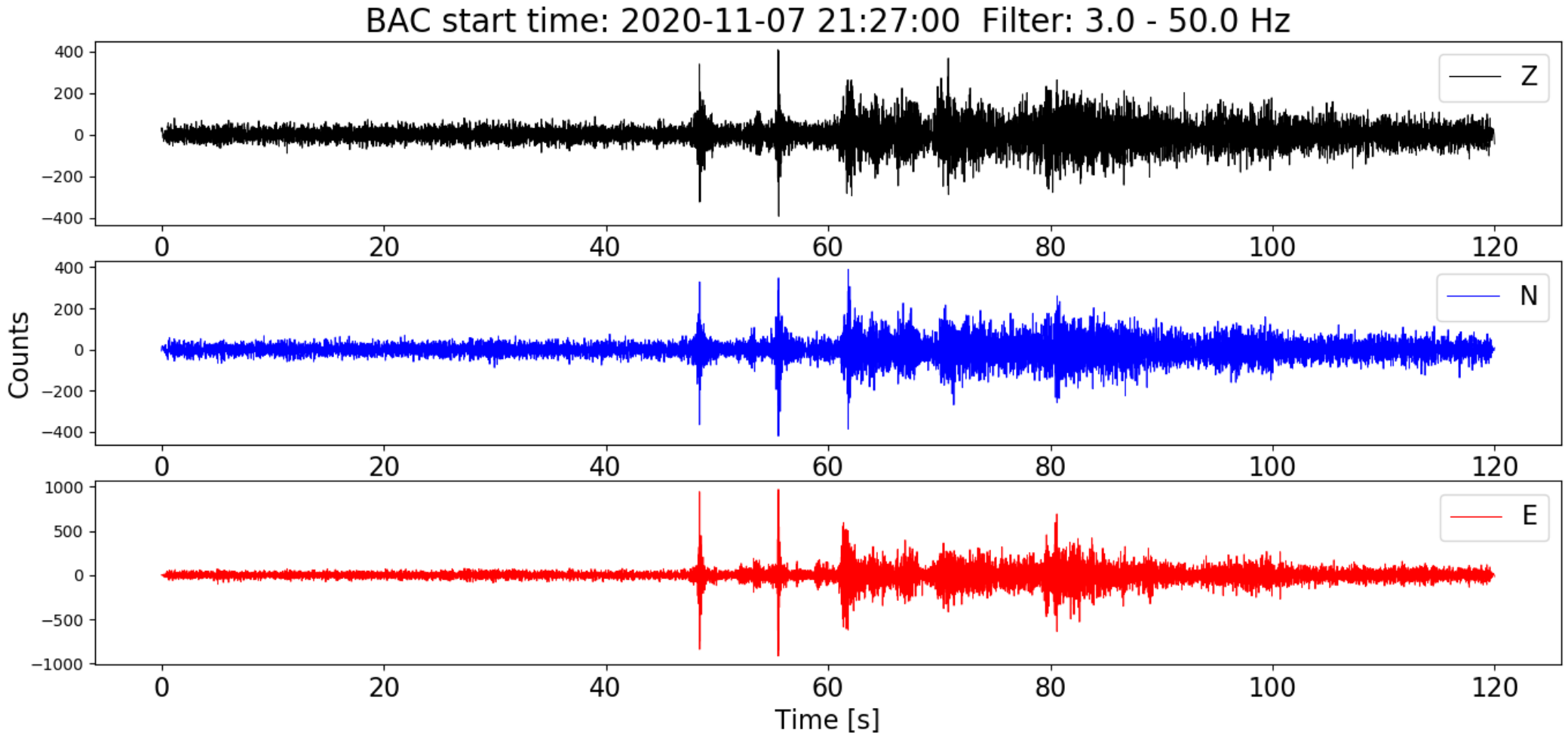

Figure 14. Audio data from Vittinge (BACU) seismic station (SNSN 2020).

We examined the arrival times of the acoustic signals at each seismometer. By focusing on two stations (OSTU and ESKU) located in nearly opposite directions from the fall site, we calculated an average speed of sound from the source through the air, arriving at approximately 295 m $s^{-1}$. This value is consistent with the reference speed of sound at altitudes between 38.9 km and 12.5 km, which ranges from 286.5 to 296 m $s^{-1}$.

Using the calculated distances from each station to the wave source as radii, we constructed spherical wavefronts around each of the five seismic stations. These spheres intersect at a common location, allowing to approximate the coordinates of the sound origin through spatial triangulation. However, one sphere, centered on the closest station (BACU), does not intersect with the others at this point. Being the nearest station, BACU likely received sound waves that travelled predominantly through warmer air layers, in contrast to the waves reaching the more distant stations. The sound waves to these further stations travelled through colder air layers for longer, effectively lowering the mean speed of sound along those paths.

Although we assume that the video frame rate and sound timing are reasonably accurate, we cannot fully dismiss the possibility of imprecision in the video timing (Table 6, Fig. 4). Thus, further refinement or additional data might be necessary to achieve a more precise match to the trajectory.

Data from both recording locations in Figure 4 and in Figure 14 distinctly shows three of the loudest sonic booms, which we attribute to three separate meteorite fragments. Regarding additional recorded booms, we consider the possibility that they could be caused by other individual fragments or, alternatively, echoes of the loudest sonic booms.

If a meteoroid survives the luminous phase of its flight (M. Gritsevich et al. 2012; E. K. Sansom et al. 2019), it continues traveling at supersonic speeds for a considerable time after the visible meteor trail has ended (J. Moilanen et al. 2021). The transition to a supersonic flight for a fragment occurs much lower than observed or calculated terminal height of the luminous flight (M. Moreno-Ibáñez et al. 2015). A sonic boom is generated by the shock wave produced by the supersonic projectile (E. A. Silber et al. 2018). This shock wave expands outward in a conical shape from the trajectory. As the projectile slows down nearing Mach 1, the cone widens and ultimately rounds off at the end of this shock wave cone. This is why the first audible sonic boom is usually caused by the shock wave of the fragment, which goes subsonic at the lowest altitude. This phenomenon is particularly pronounced in the Swedish fireball case due to the steep entry angle. For meteorite falls with shallower entry angles, however, a more detailed sound propagation model would be required to accurately determine sonic boom origins.

We estimated the sound travel times from the point where a simulated fragment goes subsonic to the two closest recording places (Björkbacken and BACU). To enhance accuracy, we calculated the speed of sound for each altitude to account for variations in air temperature and density.

Sonic-boom-filtered DFMC simulations indicate that the first loud sonic boom most likely originated from a fragment decelerating to subsonic speed at an altitude of approximately 8.7 km. This result is also consistent with the reported low-frequency "thud" when interpreted as the first sonic boom coupling into the ground and propagating as a seismic wave. The second loud sonic boom was likely generated at an altitude of approximately 9.9 km, and the third at approximately 10.6 km.

## 5 Discussion

The recovery of the iron meteorite near Ådalen marked a significant milestone as the first instrumentally recorded and confirmed iron meteorite fall, enabling the calculation of its Solar System orbit (I. Kyrylenko et al. 2023). This achievement is particularly noteworthy given the rarity of such occurrences and the prior absence of calculated orbit for iron meteorites. The analysis of this event enhances our understanding of meteor dynamics in the atmosphere and considerably extends the range of observed impact scenarios. It also provides a crucial benchmark for future studies of meteorite falls of similar origin. Although earlier studies have suggested grouping fireballs based on derived mass loss rate and deceleration characteristics (M. Gritsevich et al., 2012; E. K. Sansom et al., 2019; M. Moreno-Ibáñez et al., 2020; I. Boaca et al., 2022; Peña-Asensio and Gritsevich, M., 2026; T. Stevenson et al., 2026), this observation provides the first true data point corresponding specifically to iron material, ablation, and bulk density.

The DFMC model predicts the strewn field by propagating a large Monte Carlo ensemble of fragment trajectories that sample uncertainties in the trajectory solution, fragment properties, aerodynamic behaviour, and atmospheric conditions. Consequently, the predicted strewn field should be interpreted probabilistically rather than deterministically. The highest-density regions represent the most likely landing locations, whereas the outer regions reflect progressively less probable outcomes arising from the combined effects of observational uncertainty, atmospheric variability, and fragment dynamics during dark flight.

At the time of this study, the only publicly available indication of a possible meteorite recovery area was the updated search region published by Eric Stempels (pers. comm., 2021). This search region overlaps our predicted strewn field, although it is considerably larger and would require substantial effort to investigate thoroughly. This naturally raises two questions: how well does the recovered 13.8 kg iron meteorite agree with our modelled strewn field, and why do we not simply adjust the model to match the proposed search area? The answer is that a single recovered meteorite does not define the full extent of a strewn field. Additional recovered fragments are required to constrain its true spatial distribution.

A further challenge is the highly irregular, yet aerodynamically sculpted, shape of the recovered iron meteorite (Fig. 7, see also Supplementary Material). Such asymmetric bodies exhibit complex aerodynamic behaviour during dark flight, making their trajectories difficult to model accurately. At present, no validated models exist that can reliably predict the dark-flight behavior of meteorite fragments with this degree of shape asymmetry. Moreover, the commonly adopted values for the dimensionless product of the pre-atmospheric shape factor and drag coefficient (CdAe), typically ranging from 1.5 to 1.8 for realistically complex bodies in meteor trajectory interpretation models (e.g., H. A. R. Devillepoix et al., 2018; H. A. R. Devillepoix, E. K. Sansom, et al., 2022; Gritsevich, 2008; Gritsevich et al., 2017; M. M. M. Meier et al., 2017; P. M. Shober et al., 2022), would not be appropriate in this particular case. Refining existing models based solely on this single irregular fragment is premature and scientifically imprudent. Collecting additional fragments is essential to improving strewn field predictions and providing a more accurate representation of the meteorite fall area.

Knowing the recovery location of the 13.8 kg iron meteorite, we consider it unlikely that this specimen represents the central part of the strewn field. This interpretation is supported by the extensive search efforts conducted in the area (some of which have been documented online) and by the fact that iron meteorites are generally much easier to recognize and recover than stony meteorites. Nevertheless, no additional meteorite fragments have been reported from the vicinity of the recovery site. This raises the possibility that a substantial fraction of the search effort was concentrated outside the main strewn field.

Furthermore, there is limited knowledge regarding the depth to which iron meteorite fragments might penetrate the terrain upon impact, with only a few well-documented cases, such as Sikhote-Alin, offering insights. The situation in this case is even more unique, as the fragment impacted a boulder, which influenced both its final resting location and the visibility of potential impact evidence. Based on our analysis and simulations, it remains plausible that the meteorite struck the ground near the boulder, potentially causing a repeated ricochet effect. The nearby ditch could also have resulted from this impact, though without conclusive evidence, this remains speculative. Considering the available data, we favor the scenario in which the meteorite ricocheted off the boulder and landed at its final resting position (Fig. 10). For some fireball events even lower terminal heights have been witnessed or can be computed (M. Gritsevich & Popelenskaya, 2008, M. Moreno-Ibáñez et al. 2015, 2017), examples being Sikhote-Alin and Sterlitamak (both iron meteorites, though these were not instrumentally documented). Among well-documented cases, even the massive Chelyabinsk meteorite had a higher terminal height of 13.6 km than Ådalen, primarily due to more intense mass loss and considerably shallower entry angle of about 17°.

In the case of Ådalen, the steep trajectory allowed the largest meteorite fragments to maintain supersonic speeds down to a very low altitude. During the Chelyabinsk event, the shock wave was extremely powerful, reaching the ground 2 minutes 23 seconds after the end of the fireball. Here, however, video recordings show that the sonic booms were heard in under 29 seconds after the light caused by the fireball disappeared, reflecting the rapid atmospheric descent and proximity of the event.

The current DFMC model (J. Moilanen et al. 2021; M. Gritsevich et al. 2024) implementation of aerodynamic forces, even when applied to arbitrarily shaped fragments, does not account for lift forces. Lift could potentially cause the fragment to drift in a particular direction, especially for highly asymmetrical shapes like the 13.8 kg meteorite. Accurately incorporating a lift would require specifying the fragment shape, which would limit the model ability to deal with arbitrary shapes. This limitation means that while drag is considered, the influence of lift, which could significantly alter the trajectory of an irregularly shaped and/or rapidly rotating symmetric fragment, is not factored into the simulation, impacting the accuracy of the predicted path in such exotic scenarios.

According to photographs, the 13.8 kg meteorite has an asymmetrical shape, characterized by a flat plane with one thicker end. This irregular geometry may cause significant drift deviations from the predicted trajectory. Additionally, the light curve (Fig. 3) generated from the security video suggests a rotation period of approximately 0.6 seconds at the onset of the event. This rapid rotation could generate forces capable of diverting the meteoroid from its original path during its peak brightness. Given its shape, it is also possible that the meteorite impacted the boulder from a different direction and angle than our simulations predict.

There was no catastrophic fragmentation episodes observed, as indicated by the absence of notable flares. The light curve from the surveillance camera video in Figure 3 shows only two flares—one at 0.4 seconds and another at 1.0 second after the maximum brightness (t = 0.0 s). This pattern supports the idea of three large fragments on the ground, as mentioned in section 3.2.

The fireball of 7 November 2020 stands out as one of the most straightforward meteor events to work with due to its steep entry, unprecedentedly low penetration in the atmosphere, well-calibrated multi-station observations, sound recordings, video footage, and the recovery of readily identifiable iron meteorite fragment on the ground. Despite extensive searches of the strewn field, additional meteorite fragments may still remain undiscovered. In particular, our own searches were conducted without metal detectors because obtaining the required permits was time-consuming, thereby limiting the effectiveness of the field campaign. Moreover, even iron meteorites can be overlooked during detector surveys if the search coil does not pass sufficiently close to a buried fragment or if fragments have become deeply embedded in soft soil upon impact.

# 6 Conclusions

Based on the detailed study of the 7 November 2020 iron meteorite fall near Ådalen, Sweden, we provided a comprehensive analysis of the event and its implications. The fireball was extensively observed across Sweden, Norway, Denmark, and Finland, with sightings from distances up to approximately 665 km. The observed deep atmospheric penetration, with a terminal altitude of 11.28 km, sets the lowest documented fireball descent. This unique entry profile strongly suggested an iron composition, a hypothesis confirmed by subsequent observations and composition analyses.

The recovery and appearance of a 13.8 kg meteorite fragment, coupled with the discovery of fusion crust fragments near an impact-marked boulder, further corroborated its iron meteorite origin. These findings offered valuable insights into the meteorite's ground impact, providing a basis for further analysis of the event. Based on our *in-situ* studies and corresponding observations, we conducted a detailed investigation of a possible ricochet scenario. We tested the plausibility of the meteorite striking a boulder and ricocheting 75 meters to its final resting place, which appears to be a likely case. This is consistent with the available data and simulations, which point to this as the most plausible explanation for the meteorite's final resting position.

A major factor contributing to the distinct behavior of this event is the durability of iron meteoroids, which often retain distinct streamlined shapes with well-developed regmaglypts. These characteristics, which influence aerodynamic behavior, have not been previously considered in fireball modelling. Our study highlights that modelling iron meteoroid entry may necessitate not only different considerations of bulk density but also the incorporation of specific shape and drag effects, which can significantly alter their trajectory and descent dynamics.

The wealth of data from multi-station observations, sound recordings, and video footage enabled us to conduct detailed reconstruction of the meteor trajectory and behavior, as well as to define a potential strewn field for suggested search areas. The study of this fall has enriched the field by providing an unprecedented set of data, with implications for both the understanding of meteorite dynamics and the refinement of future meteorite search strategies. The meteorite's recovery marked a significant milestone as the first instrumentally recorded and confirmed iron meteorite fall, enabling the calculation of its Solar System orbit. This achievement is particularly noteworthy given the rarity of such occurrences and the prior absence of calculated orbits for iron meteorites.

## 7 Data availability

The original observations are publicly available through the Taivaanvahti observation database using the observation identifiers listed in Table 2. Observations from the key stations used in our study are included in the Supplementary Material. The 3D shape model of the recovered meteorite is also included in the Supplementary Material.

## 8 Acknowledgements

This work was initiated with the support of the Finnish Geospatial Research Institute and the Academy of Finland project no. 325806 (PlanetS). MG acknowledges the Spanish Ministry of Science, Innovation and Universities project No PID2023-151905OB-I00. The program of development within Priority-2030 is acknowledged for supporting the research at UrFU. We thank all members of the Finnish Fireball Network and other Finnish observers, especially Esko Lyytinen, Jari Pyykkö, Arto Oksanen, Harri Kiiskinen, Jari Tuukkanen, Markku Lintinen, Veijo Lahola, Heikki Helinkö, and Kari A. Kuure, for their dedicated commitment, and acknowledge Ursa Astronomical Association for the support with the Network coordination. We are grateful to all the other scientists and observers, who contributed to the discussions and acquisition of the data, especially Eric Stempels, Steinar Midtskogen, Johan Kero, Daniel Kastinen, Alberto J. Castro-Tirado, Michael Schieschke, and Björn Lund, as well as to geologists Andreas Forsberg and Anders Zetterqvist who reported the meteorite find, and to Johan Benzelstierna von Engeström for his enthusiasm and overall support of the meteorite searches in the area.

## ORCID iDs

Jarmo Moilanen https://orcid.org/0000-0002-6410-3709
Maria Gritsevich https://orcid.org/0000-0003-4268-6277